\documentclass[10pt]{emulateapj}
\usepackage{apjfonts}
\usepackage{psfig}
\usepackage{epsfig}

\newcommand\puncspace{\ifmmode\,\else{\ifcat.\C{\if.\C\else\if,\C\else\if?\C\else%
\if:\C\else\if;\C\else\if-\C\else\if)\C\else\if/\C\else\if]\C\else\if'\C%
\else\space\fi\fi\fi\fi\fi\fi\fi\fi\fi\fi}%
\else\if\empty\C\else\if\space\C\else\space\fi\fi\fi}\fi}
\newcommand\SP{\let\\=\empty\futurelet\C\puncspace}
\font\smfont=cmr9
\newcommand\Halpha{{H$\alpha$}\SP}                     
\newcommand\I{\kern.2em{\smfont I}\SP}                 
\newcommand\II{\kern.2em{\smfont II}\SP}               
\newcommand\fnii{\mbox{\rm [N\II]}\SP}                 
\newcommand\fsii{\mbox{\rm [S\II]}\SP}                 
\newcommand\HI{\mbox{\rm H\I}\SP}                      
\newcommand\HII{\mbox{\rm H\II}\SP}                    
\newcommand\Ib  {$I$-band\SP}                          
\newcommand\Bb  {$B$-band\SP}                          
\newcommand\Rb  {$R$-band\SP}                          
\newcommand\cf{\textit{cf.} }                          
\newcommand\eg{\textit{e.g.}, }                        
\newcommand\etal{\textit{et al.} }                     
\newcommand\ie{\textit{i.e.} }                         
\newcommand\pone  {Paper I}                            
\newcommand\ptwo  {Paper II}                           
\newcommand\pthree{Paper III}                          
\newcommand\mls {$\left[\frac{M}{L}\right]_{\star}$\SP}
\newcommand\mlb {$\left[\frac{M}{L}\right]_b$\SP}      
\newcommand\mmlb{\left[\frac{M}{L}\right]_b}           
\newcommand\mld {$\left[\frac{M}{L}\right]_d$\SP}      
\newcommand\mmld{\left[\frac{M}{L}\right]_d}           
\newcommand\kms {km~s$^{-1}$\SP}                       
\newcommand\hMpc{h$^{-1}$~Mpc\SP}                      
\newcommand\hkpc{h$^{-1}$~kpc\SP}                      
\newcommand\dV  {$\Delta V_{circ}$\SP}                 
\newcommand\Rd  {R$_d$\SP}                             
\newcommand\RI  {R$_{23.5}$\SP}                        
\newcommand\RB  {R$_{25}$\SP}                          
\newcommand\RCf {R$_{f}$\SP}                           
\newcommand\NGC {NGC\kern.33em}                        
\newcommand\UGC {UGC\kern.33em}                        
\newcommand\AGC {AGC\kern.33em}                        
\newcommand\CGCG{CGCG\kern.33em}                       
\newcommand\IC  {IC\kern.33em}                         
\newcommand\QOS {$Q_{sh}$}                             
\newcommand\QO  {$Q_{opt}$}                            
\newcommand\QH  {$Q_{{\rm HI}}$}                       
\newcommand\QI  {$Q_{im}$}                             
\newcommand\aas {A\&AS}                                

\shorttitle{Properties of Cluster Spirals I.}
\shortauthors{Vogt et al.}

\slugcomment{Accepted by AJ 2004 February 23}

\begin{document} 

\title{ M/L, H$\alpha$ Rotation Curves, and H\I Measurements     \break
        for 329 Nearby Cluster and Field Spirals:  I. Data  }

\author{Nicole P. Vogt\altaffilmark{1,2}}
\affil{Department of Astronomy, New Mexico State University, Las Cruces, NM 88003}
\email{nicole@nmsu.edu}
\and
\author{Martha P. Haynes,\altaffilmark{3} Terry Herter, and Riccardo Giovanelli,\altaffilmark{3}}
\affil{Center for Radiophysics and Space Research, Cornell University, 
Ithaca, NY 14853}
\email{haynes, riccardo, and herter@astro.cornell.edu}

\altaffiltext{1}{Formerly at: Institute of Astronomy, University of Cambridge, Cambridge, 
CB3$-$0HA, UK}
\altaffiltext{2}{Formerly at: Center for Radiophysics and Space Research, 
Cornell University, Ithaca, NY 14853}
\altaffiltext{3}{National Astronomy and Ionosphere Center;  NAIC is operated 
by Cornell University under a cooperative agreement with the National Science 
Foundation.}

\begin{abstract}
A survey of 329 nearby galaxies (redshift z $<$ 0.045) has been conducted to
study the distribution of mass and light within spiral galaxies over a range
of environments.  The 18 observed clusters and groups span a range of
richness, density, and X-ray temperature, and are supplemented by a set of 30
isolated field galaxies.  Optical spectroscopy taken with the 200-inch Hale
Telescope provides separately resolved \Halpha and \fnii major axis rotation
curves for the complete set of galaxies, which are analyzed to yield velocity
widths and profile shapes, extents and gradients.  \HI line profiles provide
an independent velocity width measurement and a measure of \HI gas mass and
distribution.  \Ib images are used to deconvolve profiles into disk and bulge
components, to determine global luminosities and ellipticities, and to check
morphological classification.  These data are combined to form a unified data
set ideal for the study of the effects of environment upon galaxy evolution.
\end{abstract}


\section{Introduction}
\label{ch:intro}

This study concerns the effects of the cluster environment on the distribution
of mass and light in spiral galaxies.  These data are part of a large survey
of nearby clusters (redshift $z \le 0.045$) designed to study the dependence
of mass--to--light distributions upon cluster densities by comparing core
galaxies to periphery and low density region members, to evaluate the global
properties of rotation curves, and to examine fundamental galaxy properties in
varied environments.  There are several other large samples of optical
rotation curves in the literature at present, including Courteau's 1992 sample
of 350 which focuses primarily upon field galaxies, and Mathewson \& Ford's
1996 sample of 2447 in the southern hemisphere.  This data set contains the
rotation curves of 329 northern hemisphere cluster and field galaxies, with
\HI observations for most to provide additional, independent constraints upon
mass distributions.

The velocity profiles of spiral galaxies were originally believed to fall in a
Keplerian decline at large radii, following an inverse $r$ law near the edge
of the optical disk where the light (and presumably mass) dropped off.  When
the first measured rotation curves did not show such a trend they were assumed
to probe only the inner region, rather than that predicted to exist at large
radii for a pure exponential disk model.  As hardware improvements extended
the sensitivity and depth of observations, however, more and more of the outer
regions of galaxy disks were observed.  It became clear that both optical and
21~cm line observations showed velocity profiles which rose in the inner
regions but often remained flat at their outer extents, implying the presence
of a considerable amount of dark matter within the optical radius (\cf Bosma
1978, review by Faber \& Gallagher 1979).

In the early 1980's Rubin and collaborators began a pioneering study of spiral
galaxies to attempt to constrain mass distributions by analyzing optical
velocity profiles and light distributions (Rubin, Burstein, Ford, \& Thonnard
1985, and references therein).  They deduced that velocity profiles correlated
well with basic galaxy properties as luminosity (brighter galaxies having
larger velocity widths which rose more slowly) and morphological type (earlier
type galaxies having larger velocity widths). This initial sample of field
galaxies was later extended to include a set of cluster spirals (Burstein,
Rubin, Ford, \& Whitmore 1986; Rubin, Ford, \& Whitmore 1988; Whitmore,
Forbes, \& Rubin 1988), to evaluate the importance of the cluster environment
upon the mass distributions.  They found a population of cluster galaxies with
rotation curves which appeared to decline at large radii, and argued that
these galaxies had deficient halos which had either been stripped through an
interaction with the cluster medium (\eg ram pressure sweeping) or with close
neighbors (tidal stripping), or had never been allowed to form.  Later studies
(Guhathakurta \etal 1988; Forbes \& Whitmore 1989; Distefano
\etal 1990; Amram \etal 1993; Adami \etal 1999) found conflicting results 
regarding these trends, and thus part of the motivation for this work has been
to re-examine these claims within a larger sample of galaxies.

The strong effect of the cluster environment upon the \HI gas envelope of
spiral galaxies has been well established (\cf Haynes \& Giovanelli 1986,
Magri \etal 1988, Haynes 1989, Solanes \etal 2001).  Galaxies interacting with
the hot intracluster medium of moderate to high X-ray luminosity clusters are
observed to be strongly \HI deficient and to have lost \HI gas preferentially
at outer radii (Haynes 1989), and the pattern of \HI deficiency correlates
strongly with clustercentric radius.  The stripped galaxies appear to have an
undisturbed spiral morphology, though biased towards early types, implying
that the mechanisms in play are either too weak to directly affect the stellar
material or that the effect can only be recognized on a longer timescale (\eg
decreased young star formation from a depleted \HI gas reservoir).

This paper is the first of three in a series.  The present paper (\pone)
details the observations which were made in the optical and in the radio, and
defines the data analysis procedures and the modeling technique used to
determine mass-to-light ratios\ for the galaxies within the sample.  The second paper (Vogt
\etal 2004a; \ptwo) investigates the evidence for galaxies which are currently
infalling into the cores of clusters on a first pass, or show evidence of a
previous passage through the core, and the role of gas stripping mechanisms in
a morphological transformation of the field spiral population into cluster
S0s.  The third paper (Vogt \etal 2004b; \pthree) explores differences
in fundamental galaxy properties (size, mass, and luminosity) as a function of
environment, whether spiral disks within rich cluster cores may have coalesced
from their halos at an early epoch, and examines the relationship between \HI
gas stripping by the hot intracluster medium and the consequential suppression
of young star formation across the disks.  

\section{Sample Properties}
\label{sec:smpprp}

\subsection{Distribution of Clusters}
\label{sec:smpcls}

\begin{figure} [htbp]
  \begin{center}\epsfig{file=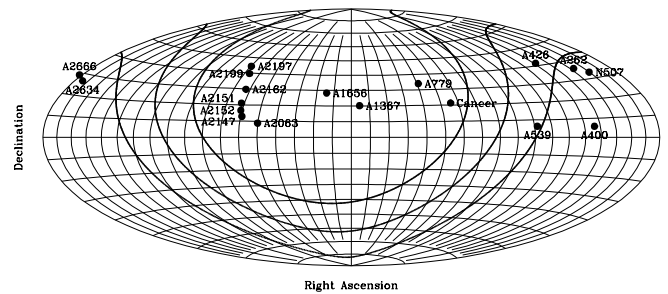,width=3.4truein}\end{center}
  \caption[Cluster Distribution]
  {The cluster sample shown on an Aitoff equal area projection of the
  celestial sphere, centered at 12 hours.  Heavy lines define the galactic
  plane and $\pm 20^{\circ}$ about it.}
  \label{fig:skyplt}
\end{figure}

We have assembled herein the data from an observational program in moderate
resolution optical spectroscopy conducted at the 200-inch Hale Telescope, with
parallel 21~cm and \Ib photometric observations, centered upon a set of 18
nearby clusters and an accompanying field sample.  The clusters span a wide
range in richness, density, and X-ray temperature in order to explore a wide
range of environments, with redshifts between 5000 \kms and 12000 \kms.  They
were selected so that part of the sample was overhead at all hours, to enable
observations at all seasons and maximize our use of observing time on the
project, with a preference for clusters within the Arecibo Observatory
declination range ($-1{^\circ} < \delta < 38^{\circ}$).  The initial selection
of 16 was made from the northern catalog of rich clusters of galaxies (Abell,
Corwin, \& Olowin 1989), drawing from the Pisces--Perseus, Hercules, and Coma
superclusters.  We began with six rich clusters for which many \HI profiles
had been already obtained and X-ray flux data existed (see Magri \etal 1988
study of \HI deficiency): A262, A426 (Perseus), A1367, A1656 (Coma), A2147,
and A2151 (Hercules).  We added A2152 as it overlaps with A2147 and A2151, and
A2197 and A2199 as the companion clusters cover the extremes of rich (A2199)
and poor (A2197) environments.  More moderate environments were sampled within
clusters A400, A539, and A2063, and A2634 and close neighbor A2666, and the
poor clusters A779 (chosen also to fill a gap at $\alpha \sim 10^h$) and A2162
(selected but observations barely begun).  The sample was then augmented by
the Cancer (\CGCG0819.6+2209) and \NGC507 (\CGCG0150.8+3615) groups.  The
cluster distributions were evaluated after the observations were completed and
membership criterion imposed to separate true cluster members from those
associated with the cluster in the nearby supercluster envelope and foreground
and background field galaxies (see discussion in \ptwo).
Table~\ref{tab:cl_lst} lists the cluster positions, redshifts, and parent
supercluster structures, and Figure~\ref{fig:skyplt} shows their distribution
upon the sky.  Redshifts are presented in the heliocentric and the CMB
reference frames (see Kogut \etal 1993 for relative relations).  A number of
these clusters were analyzed in the peculiar velocity study of Giovanelli
\etal 1997; coordinates and redshifts are consistent between the two
presentations.

\begin{table*} [htbp]
  \caption{Distribution of Clusters}
  \begin{center}
  \begin{tabular} {l c l r r r l @{\hspace{-1.5em}} r} 
  \tableline
  \tableline
  Cluster   & \multicolumn{2}{c}{R.A. $\;\;(1950)\;\;$ Dec.} & \multicolumn{2}{c}{Redshift$^a$} & $R_A^b$ & Supercluster & References \\
  \multicolumn{1}{c}{(1)}  & \multicolumn{1}{c}{(2)}  & \multicolumn{1}{c}{(3)}  & \multicolumn{1}{c}{(4)}  & 
  \multicolumn{1}{c}{(5)}  & \multicolumn{1}{c}{(6)}  & \multicolumn{1}{c}{(7)}  & \multicolumn{1}{r}{(8)} \\
  \tableline 
  N507      & 01 20 00.0 & +33 04 00 &  5091 &  4808 & 1.74 & Pisces-Perseus      & 1    \\
  A262      & 01 49 50.0 & +35 54 40 &  4918 &  4664 & 1.81 & Pisces-Perseus      & 1    \\
  A400      & 02 55 00.0 & +05 50 00 &  7142 &  6934 & 1.25 & Pisces-Perseus      & 2    \\
  A426      & 03 16 20.0 & +41 20 00 &  5460 &  5300 & 1.63 & Pisces-Perseus      & 3    \\
  A539      & 05 13 54.0 & +06 25 00 &  8730 &  8732 & 1.04 & \nodata             & 4    \\
  Cancer    & 08 17 30.0 & +21 14 00 &  4705 &  4939 & 1.91 & \nodata             & 5    \\
  A779      & 09 16 48.0 & +33 59 00 &  6967 &  7211 & 1.30 & \nodata             & 2    \\
  A1367     & 11 41 54.0 & +20 07 00 &  6408 &  6735 & 1.38 & Coma                & 2    \\
  A1656     & 12 57 24.0 & +28 15 00 &  6917 &  7185 & 1.29 & Coma                & 6    \\
  A2063     & 15 20 36.0 & +08 49 00 & 10445 & 10605 & 0.86 & Hercules, southern  & 2    \\
  A2147     & 16 00 00.0 & +16 02 00 & 10493 & 10588 & 0.87 & Hercules, canonical & 7    \\
  A2152     & 16 03 07.0 & +16 35 00 & 12930 & 13018 & 0.72 & Hercules, canonical & 7    \\
  A2151     & 16 03 00.0 & +17 53 00 & 11005 & 11093 & 0.83 & Hercules, canonical & 7,8  \\
  A2162     & 16 10 30.0 & +29 40 00 &  9600 &  9659 & 0.95 & Hercules, northern  & 9    \\
  A2197     & 16 26 30.0 & +41 01 00 &  9138 &  9162 & 0.99 & Hercules, northern  & 2,10 \\
  A2199     & 16 26 54.0 & +39 38 00 &  8970 &  8996 & 1.00 & Hercules, northern  & 2,10 \\
  A2634     & 23 35 54.9 & +26 44 19 &  9240 &  8895 & 0.98 & Pisces-Perseus      & 11   \\
  A2666     & 23 48 24.0 & +26 48 24 &  8118 &  7776 & 1.11 & Pisces-Perseus      & 2,11 \\
  \tableline \\
  \multicolumn{8}{l}{\hspace{0.05truein} $^a$Heliocentric and CMB reference frames.} \\
  \multicolumn{8}{l}{\hspace{0.05truein} $^b$Abell radius (degrees).} \\
  \multicolumn{8}{l}{References. --- (1) Sakai, Giovanelli, \& Wegner (1994), (2) Zabludoff \etal }  \\
  \multicolumn{8}{l}{(1993a), (3) Kent \& Sargent (1983),  (4) Ostriker \etal (1988), (5) Bothun }   \\
  \multicolumn{8}{l}{\etal (1983), (6) Kent \& Gunn (1982), (7) Barmby \& Huchra (1997), (8) Bird, } \\
  \multicolumn{8}{l}{Dickey, \& Salpeter (1993), (9) Abell (1958), (10) Dixon, Godwin, \& Peach } \\
  \multicolumn{8}{l}{(1989), (11) Scodeggio \etal (1995).} \\
  \end{tabular}
  \end{center}
  \label{tab:cl_lst}
\end{table*}

We began by characterizing the clusters by a set of criteria, including the
Abell richness class, the diffuse X-ray gas luminosity and temperature, the
velocity dispersion, and the \HI deficiency of member galaxies.  The Abell
classification serves as a robust measure of cluster richness, defined by the
number of bright galaxies within one $R_A$ (1.5 \hMpc).  Rich clusters are
also distinguished by the presence of a large, diffuse envelope of hot X-ray
gas, of order $10^{42}$ to $10^{45}$ ergs $s^{-1}$ (Beers \etal 1991), with a
flux distribution only roughly equivalent in some cases to that of the
individual galaxies (the luminous matter).  This material can comprise up to
30\% of the mass of a cluster (Dell'Antonio, Geller, \& Fabricant 1995).  It
is accreted from the surrounding 10 -- 20 Mpc (David 1997), in contrast to the
dark matter, which can be accounted for entirely by the stripping of
individual galaxy halos.

Dynamically evolved, centrally condensed clusters such as A1656 have high
X-ray luminosities and intergalactic gas temperatures and the X-ray flux is
distributed smoothly throughout the cluster.  In contrast, less evolved rich
clusters such as A2634 have lower X-ray luminosities and temperatures and the
X-ray intergalactic gas is clumped about individual galaxies, suggesting
recent emission or gas stripping by ram pressure (Gunn \& Gott 1972) or tidal
forces (\cf Toomre \& Toomre 1972, and more recently Weinberg 1996).  The
X-ray gas distribution and temperature serves as an indicator of the overall
state of the cluster: the X-ray temperature correlates tightly with the
velocity dispersion (\ie the depth of the potential well), while the X-ray
luminosity reflects the density of the intracluster medium.

\HI deficiency, in contrast, is as a specific indicator of the {\it local} 
environment around a particular galaxy.  By observing the morphology-density
relation (\cf Dressler 1980a), number counts of galaxies, and \HI deficiency
statistics, we gain a measure of the previous interaction history of member
galaxies, with both the intergalactic medium and with other galaxies within
the potential.  We are thus able to characterize the range of our sample, from
such highly evolved clusters as A1656 and A426, to the loose groups of Cancer
and \NGC507, to isolated field galaxies.

\subsection{Galaxy Selection}
\label{sec:smpgal}

An initial search was made of the Upsala General Catalog (UGC, Nilson 1973)
for sources which fell within several $R_A$ of the target cluster centers (the
total sample is heavily concentrated within one $R_A$).  Extensive use was
made of the source lists contained within Dressler's catalog (1980) of rich
clusters, and the accumulated sources already entered into the 1994 version of
the private database of R.G. and M.P.H. known as the AGC, containing a large number
of galaxies observed with the Arecibo Observatory 305--meter radio telescope
and/or the 300--foot NRAO telescope in Green Bank.  A visual examination was
then conducted of the POSS prints of the region surrounding each cluster
center for additional sources.  Galaxies were chosen on the basis of apparent
cluster membership (\ie within several $R_A$ of the target cluster centers,
note that redshifts were unknown for many sources before the observations
began) and spiral appearance and morphology, with no evidence of current
interaction with other sources or overlapping field stars to obscure the
profiles.  Roughly 12\% of the selected galaxies are barred systems.  The
sources were prioritized to have an angular size $\theta \ge 30$\arcsec, and
an inclination angle $i \ge 30^{\circ}\ (b/a \le 0.87)$.  The clusters were
examined from the center outwards, focusing first upon all galaxies within one
$R_A$ and then adding additional sources further out as observing time
permitted.  As the optical spectra, and \Ib images were acquired in parallel
over the five year observational period, spectral candidates were at times
prioritized within a cluster sample because an image had already been
obtained.  This was a random effect, however, as the imaging program was
designed to lag the spectral and \HI line profile programs deliberately, so as
to be comprised mainly of sources for which we had already obtained a velocity
profile.  Because of this time lag and the limited amount of observing time
available for imaging, several clusters lack substantial imaging and we have
no images for A2063.

The sample cannot be said to be strictly limited nor complete in either size
nor magnitude, due to the varying inputs into the total set.  We conducted a
program to evaluate the distances to many of the clusters via the
Tully--Fisher (Tully \& Fisher 1977) relation in parallel with this project,
and thus every effort was made to observe all spiral galaxies which appeared
to be members of the target clusters (\ie neither foreground nor background
objects, and without obscuring sources), especially in the inner regions of
the clusters.  As expected, the fraction of late type spirals decreases within
the cluster cores, though such galaxies were prioritized in the selection
process; the median type across the complete sample is Sbc.

\begin{figure} [htbp]
  \begin{center}\epsfig{file=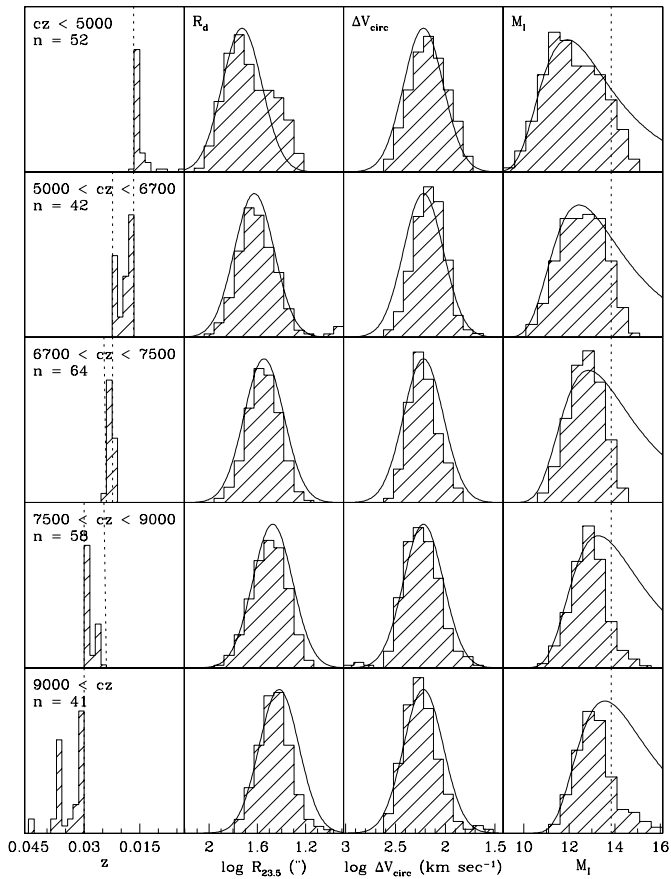,width=3.4truein}\end{center}
  \caption[Sample Selection]
  {Exploration of sample selection in the observational plane, showing the
  sample distribution in redshift, size, circular velocity, and luminosity
  broken into five redshift bands.  A Gaussian function is fit to the size and
  velocity functions, it is frozen across the redshift bins for velocity and
  shifted for those in size to account for the change in apparent size on the
  sky with redshift.  A Schechter luminosity function (M$^*$=-22.1,
  $\alpha$=-0.50, assuming the central redshift of each band) is similarly
  applied to the magnitude histograms and shifted to account for the change in
  apparent magnitude on the sky with redshift; a dotted line marks where
  completeness begins to fall off, at M$_I$ = 13.85.  The spread of each
  luminosity histogram is a combination of variations in distance and in
  intrinsic luminosity, but the first factor is minimized by the narrowness of
  the redshift bins and the strong peaks due to clustering.}
  \label{fig:sampl_sel}
\end{figure}

Figure~\ref{fig:sampl_sel} addresses the issue of sample selection, given the
lack of completeness in the selection criteria (note that the sample covers a
factor of two in redshift).  The data are divided into five narrow redshift
bins, and the distinct peaks within each bin are caused by the tight cluster
distribution.  The distribution of angular size (characterized by R$_{23.5}$,
the radius at which the disk reaches a surface brightness level of M$_I$ =
23.5 magnitudes per square arcsecond) within each bin can be fit well by a
Gaussian curve, given the narrowness of the bins (\ie the spread in distance
is rather small within each bin).  We find that a Gaussian of constant width
will fit the data across all five bins.  The mean of the curve has been
shifted to account for the change in apparent size due to distance for the
central redshift of each bin, and matches the progression in the data well.
The circular velocities are also well fit by a Gaussian, and the constant
width and mean suggest that we are sampling galaxies of the same velocity
width (mass) at all redshifts.  The change in apparent magnitudes mirrors that
in apparent size.  We fit a Schechter luminosity function to the data within
each redshift bin, a technique which works purely because the scatter in
distance within each bin is small and so the spread in luminosity is primarily
due to intrinsic variation.  The point is simply to characterize the change in
apparent magnitudes, which are well fit by a luminosity function where M$^*$
is set by assuming all galaxies within each bin fall at the central redshift
for the bin.  The completeness in magnitude can then be observed to fall off
at M$_I$ = 13.85, across all redshift bins, below the peak of the luminosity
function.

\section{Optical Spectra} 
\label{sec:rc_obs}

\subsection{Data Acquisition and Reduction} 
\label{sec:rc_data}

Moderate--resolution long--slit optical spectra were obtained for 329 spiral
galaxies with the Hale 200--inch telescope using the Double Spectrograph (Oke
\& Gunn 1982) at the Cassegrain focus ({\it f}/15.7) on 51 nights spread
throughout 12 observing runs from April 1990 to June 1994.  The red (5200 -
11000\AA) camera was used with one of two TI 800$\times$800 chips, with a 1200
l~mm$^{-1}$ grating (7100\AA\ blaze) yielding a resolution of 0.81
\AA\ pixel$^{-1}$ for 15$\mu$ pixels.  The camera has a spatial scale of
0\arcsec .58 pixel$^{-1}$ and the 1\arcsec $\times$ 125\arcsec slit was used
(only one galaxy within the sample extended beyond a slit edge in
\Halpha).  We observed the appropriately redshifted \fnii line pair ($\lambda
\lambda$6548, 6584), \Halpha ($\lambda$6563), and the \fsii\ line pair
($\lambda \lambda$6717, 6731) by obtaining spectra over the wavelength range
$6300 \le \lambda \le 7000$, with a resolution in velocity space of 37 \kms
per pixel.  Note that at a redshift $z$ = 0.045 \Halpha is redshifted to a
wavelength of $\lambda$6858, comfortably within our upper limit.

The spectrograph was rotated to align the slit along the major axis of the
source with an accuracy greater than that of the position angle measurement
($\pm2^{\circ}$, measured from sample \Ib images or Palomar Optical Sky Survey
prints).  The seeing conditions ranged from 0\arcsec .8 to 2$''$ and were
usually under 1\arcsec .5.  Because of this, neighboring 0\arcsec .58 pixels
are not independent.  Sources were observed on average at less than 1.2
airmasses, and never at more than 1.5 airmasses.  Exposures ranged from 2400
to 3000 seconds in length.  As the project progressed, integration times were
lowered slightly from the nominal value of 3000 seconds for all but the most
distant clusters in order to observe more sources per night, as the S/N ratios
were quite high in the observed emission lines.

The data were reduced using standard Image Reduction and Analysis\footnote{IRAF
is distributed by the National Optical Astronomy Observatories, which are
operated by the Association of Universities for Research in Astronomy (AURA)
under cooperative agreement with the National Science Foundation.} (IRAF)
routines and custom software.  The CCD was aligned so that the spectra were
dispersed along the columns, and the spatial axis ran along the rows.  Flat
fields were created from dome flats taken with high intensity white light
lamps, and standard stars were placed in a series of five positions across the
slit twice each night to correct for S--curve distortion along the chip (0.1 to
0.2 pixels).  Spectra were wavelength calibrated directly from the strong night
sky emission lines as they spanned the wavelength range observed quite well,
and to avoid the problems of variable spectrograph flexure which appeared in
the lamp spectra taken directly before and after each exposure.  

A region was marked on each spectrum beyond the extent of the disk emission
(for all but the most spatially extended galaxies), and used to calculate an
average background level across the chip. This background was then subtracted,
to remove the strong night sky line emission which fell near or on top of the
galaxy emission lines.  A running (boxcar) average was applied across the chip
in a box three columns wide, to smooth the data slightly.  A Gaussian fitting
routine was then applied to each column of data in the zone about the \Halpha
line, fit in either emission or in absorption, and then the stronger \fnii
line ($\lambda$ 6584) to trace the velocity profile across the optical disk.
The \fsii\ line pair ($\lambda \lambda$6717, 6731) had been redshifted into a
wavelength range which is heavily contaminated by night sky lines for most of
the galaxies in this sample, and thus the lines could not be well fit.
Criteria were set for the quality of the fit at every column: an amplitude
requirement (5$\sigma$), a width requirement (3$\sigma$), a limit on the line
width (1 - 10\AA), and a limit on the distance that the line centerpoint could
shift between two columns (5 pixels).  This fitting process was iterated until
the fit of the lines was found to be adequate.  The final fits are good to
within 5 \kms.

\subsection{Comparison with Literature Data} 
\label{sec:rc_lit}

\begin{figure} [htbp]
  \begin{center}\epsfig{file=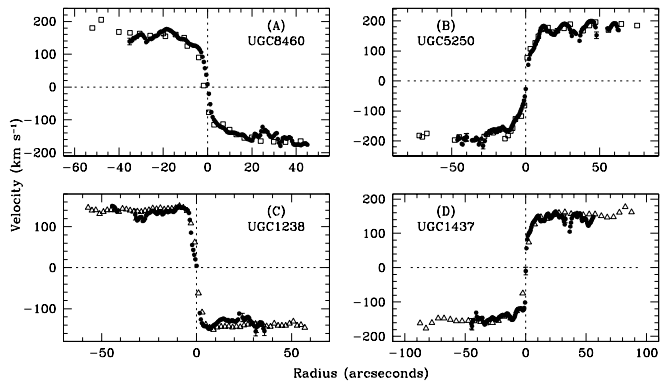,width=3.4truein}\end{center}
  \caption[Comparison Rotation Curves]
  {Rotation curves from our sample and from the literature for overlapping
  sources, plotted on top of each other for comparison.  The \Halpha profiles
  are uncorrected for inclination and in the rest frame of the galaxy; small
  circles represent our data, squares (A and B) the data of Rubin \etal (1982,
  1980), and triangles (C and D) the Fabry--P\'{e}rot data of Amram \etal
  (1994).  As Amram \etal (1994) collapse their two--dimensional velocity maps
  to a radial profile, we have reflected this radial data about the
  centerpoint to simulate the two sides of the galaxy observed in a longslit.
  The observations agree well, though the Fabry-P\'{e}rot extends further out
  in the galaxy due to increased S/N, and shows less small scale structure due
  to azimuthal smoothing.}
  \label{fig:rc_ovl}
\end{figure}

The first significant sample of optical rotation curves was acquired by Vera
Rubin and her collaborators in the early 1980's, taken on photographic plates
with the KPNO and CTIO 160--inch telescopes.  We have re-observed six galaxies
in the Rubin sample; the spectra and derived \Halpha rotation curves are found
to be in good agreement, as one would expect given the similarity of
observational techniques.  In Figure~\ref{fig:rc_ovl}A and B we compare the
data for two common galaxies, \UGC11810 (\NGC1401) and \UGC5250 (\NGC
2998).  The overall structure is the same in both set of spectra, though as
the Rubin data has been smoothed significantly it shows less of the small
scale variations seen in our data.  Their data appear to extend slightly
further out along the galaxy disks, though of course such measurements are a
strong function of the fitting constraints applied to extracting rotation
curves, but the additional data points does not show additional structure in
the rotation curves.

The extensive data set of Mathewson \& Ford (1996) is certainly worthy of
note; however, due to its southern declination we do not overlap.  More
recently, Courteau (1992, 1997) has presented CCD longslit spectroscopy from
the Shane 120--inch of \Halpha and \fnii lines for 304 spiral galaxies.  We
have observed 23 galaxies in common; a comparison plot of the full set can be
found within Courteau 1997, and is thus not included here.  Courteau's data
were obtained primarily to determine a full velocity width for each galaxy
rather than to examine the structure of the entire rotation curve, and so are
not as deep.  Our observations have a higher S/N ratio due to longer exposures
times (50 minutes versus 25 -- 30 minutes) and a larger aperture.  Taking this
into account, we find good agreement between the two data sets.  The radial
extents of the profiles are similar, and much (though not all) of the small
scale structure is confirmed in each set.

Longslit spectroscopy observations are handicapped by the fact that they sample
only along the (major) axis of a galaxy, and thus are extremely sensitive to
the (1) detection, and (2) localized non--circular motions of individual
\HII regions or spiral arm structure {\it along this axis}.  An excellent
alternative technique is to acquire full two dimensional velocity maps via
interferometry.  Fabry--P\'{e}rot data sets typically penetrate further and
with greater S/N in the outer regions of galaxies, and full velocity maps allow a
more sophisticated deprojection of velocities into the edge--on plane by
fitting a series of disks of varying inclination as a function of radii, and
modeling for warps and substructure as a function of azimuth.  The costs,
however, in both observing and analysis time, can be high, and to date have 
limited the number of galaxies analyzed via this technique.  Amram \etal (1992, 
1994) used the 140--inch CFH telescope to obtain two--dimensional \Halpha maps 
for 36 galaxies within nearby clusters (13 of which fall within our sample), 
and the GHASP project (Garrido \etal 2002, 2003) has recently published similar 
data for 38 galaxies out of an observed sample of over 160 field spirals and 
irregulars.  Figure~\ref{fig:rc_ovl}C and D overlay the optical rotation curves 
from our data and from Amram \etal for \UGC1437 (\NGC753) and \UGC1238 
(\NGC668).  As expected the Fabry--P\'{e}rot data extends roughly 30\% further 
out in the outer regions of the galaxy, and, being smoothed azimuthally, shows 
less small scale, localized variations.  The curves agree well on global 
structure, however, and within the inner arcminute radius sampled within our 
data the terminal behavior of the velocity profile is well established.

We have compared the common galaxies within our sample and several from the
literature to establish the equivalence of techniques of observation and
analysis between various practitioners of longslit spectroscopy.  Though not
shown here explicitly, our data acquisition and reduction techniques are also
in good agreement with the published work of Dale \etal (1997, 1998, 1999),
which shares many common elements.  More extensive comparisons could be made
with a broader range of data sets (\eg Corradi \& Capaccioli 1991), but we
have shown sufficient consistency in form for our purpose.  While
Fabry--P\'{e}rot observations give additional information regarding the
detailed velocity structure of galaxy disks, the global properties obtained
via longslit work are in agreement with those derived from these measurements.

\subsection{Data Characterization and Analysis} 
\label{sec:rc_ana}

\begin{figure*} [htbp]
  \begin{center}\epsfig{file=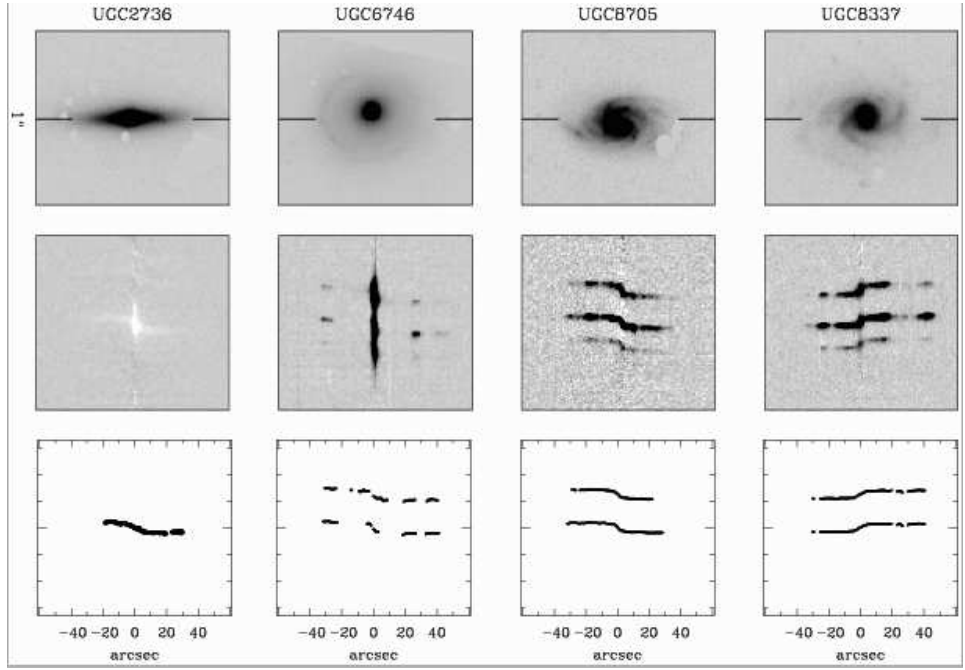,width=5.0truein}\end{center}
  \caption[Sample image and spectra]
  {We illustrate the characteristics of the sample with four galaxies; scales
  are identical and spatial axes aligned for each set of three plots.  The
  upper image shows the \Ib data for each galaxy, with the width of the
  longslit represented as a black line.  The lower image is of the optical
  spectrum centered about the \Halpha line and companion \fnii lines, with the
  bulk of the continuum removed to reveal the shape of the rotation curve in
  the inner regions.  The lower plot shows the rotation curve fit to \Halpha
  and to the stronger of the two \fnii profiles.  \UGC2736 is typed as an
  Sab, and shown because the \Halpha line is found purely in absorption,
  representative of an older stellar population; less than 10\% of the sample
  shows no \Halpha emission.  \UGC6746 one of the few Sa galaxies in the
  sample,, \UGC8705 an Sb galaxy, and \UGC8337 an Sc galaxy; these
  particular galaxies were chosen because their face-on orientation makes the
  spiral structure easier to observe in the images, but are otherwise
  representative.}
  \label{fig:plt1}
\end{figure*}

\begin{figure} [htbp]
  \begin{center}\epsfig{file=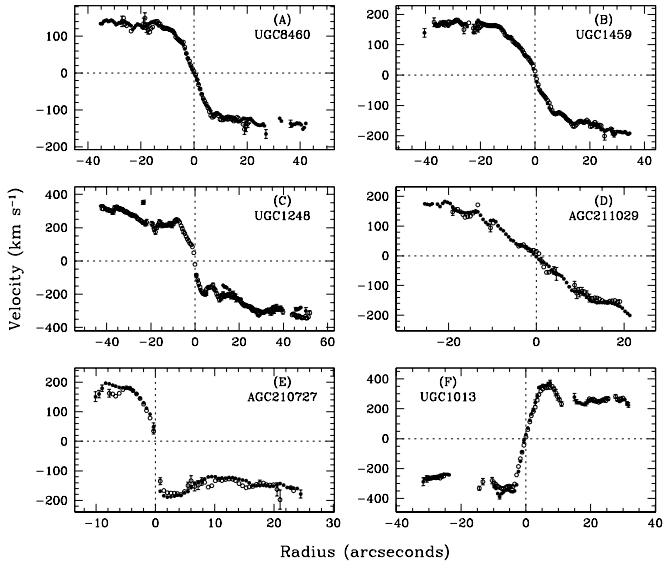,width=3.4truein}\end{center}
  \caption[Sample Rotation Curves]
  {Sample rotation curves illustrating the range of shapes within the data:
  (A) \UGC8460, with a flat \QOS = 1, (B) \UGC1459, with a slightly rising
  \QOS = 2, (C) \UGC1248 with a rising \QOS = 3, and (D)\AGC211029 with a
  solid body curve (\QOS = 4).  For (E) \AGC210727/\CGCG097-125 the two sides 
  of the rotation curve are traced to very different extents, and for (F) 
  \UGC1013 (barred) the velocity of the inner regions is higher than the 
  terminal velocity profile.  The \Halpha points are filled circles and the 
  \fnii data are open circles; error bars are plotted only when greater than 
  10 \kms.}
  \label{fig:rcplt_smp}
\end{figure}

We begin by visually characterizing the optical spectra; Figure~\ref{fig:plt1}
shows a subset of four galaxies ranging from type Sa through Sc.  These
moderate resolution data show a clear separation between the \Halpha and \fnii
lines, and the velocity profiles of the two species exhibit similar global
behavior and structure.  The \Halpha flux is generally stronger than that of
the \fnii at a given position, and tends to extend slightly further to large
radii.  Both are sensitive to the positioning of individual \HII regions along
the slit and can become patchy when traced along a late type disk with strong
spiral arms, where the regions between the arms have been swept clean of gas.
The \fnii lines tend to be better resolved in the nuclear region. The \Halpha
line, in contrast, can broaden appreciably and may be derived from clumps of
gas traveling at noncircular (\ie random) velocities.  The presence of a bar
often creates a plateau, or a local maximum, in velocity at the end of the
bar, as shown for \UGC1013 in Figure~\ref{fig:rcplt_smp}F.  The emission
extends out to a mean value of 3.7\Rd.  It can be traced past 2.2\Rd, the
point of maximum velocity for a pure exponential disk (Freeman 1970), for all
but the lowest quality data and a number of galaxies with strong bars, where
\Rd has been measured to be quite large ($> 10$ kpc).

The first galaxy in Figure~\ref{fig:plt1}, \UGC2736, is shown as it exhibits
no \fnii, and the \Halpha line is found purely in absorption.  We note that at
high inclination angles the morphological distinction between an early-type
spiral galaxy and an S0 becomes more difficult to discern, because the spectra
are composed primarily of light from the older stellar populations of the disk
(see discussion in \ptwo).  There are less than 30 such spectra in the sample.
The next three galaxies were selected because they were less inclined than
average, to allow easy examination of the spiral structure along the longslit,
but are otherwise representative of the sample (though recall that the median
type is Sbc).  The strength of the flux along the disk is a clear function of
the spiral arm structure, and indeed much of the disk of the Sa galaxy is
extremely faint.

Rotation curves were created for the entire data set by combining the \Halpha
and \fnii data together to create unified velocity profiles.  The rotation
curve of a normal spiral galaxy can be characterized by three regions.  There
is a steeply rising inner region, a transition zone (often called the elbow
point) where the rotation curve turns over, and then an outer region which is
roughly linear (like the inner region) and often flat.  We find the overall
shape of the rotation curve to be more a function of luminosity than of
morphological type, as reported by Rubin, Ford, \& Whitmore (1988), indicating
that optical morphology is not a strong measure of the underlying
gravitational potential.  The terminal velocities correlate strongly with
luminosity, as expected from the Tully--Fisher relation, and the higher
luminosity galaxies also have rotation curves which rise more quickly to their
terminal value than the lower luminosity sources (characterized by gentler
curves with wider transition zones).

The spectra were binned according to quality and to rotation curve shape.  The
278 quality (1) rotation curves are well defined (traced) and well sampled,
the 45 quality (2) curves are of fair quality; the curve is untraceable at
certain points along the disk (due either too weak \Halpha flux or to the
inference of [OH] emission lines), and the six quality (3) curves do not
adequately map the velocity profile of the galaxy (the \Halpha flux is too
weak or the galaxy is too face--on).  The shape scale is sensitive to the
slope of the rotation curve, particularly in the outer region, and defined as
follows: the 120 (1) curves are flat in their outer regions, and evenly
terminated, the 128 (2) curves are {\it almost} flat and yet {\it slightly}
rising in their outer regions, the 57 (3) curves terminate with a rising
profile, and the 12 (4) curves exhibit a pattern of solid body rotation, being
lines of one continuous slope with no turnover, or elbow, point or inflection
recorded in the emission line profile.  Examples of all four types are shown
in Figure~\ref{fig:rcplt_smp}.  Twelve of the rotation curves fit none of
these shape categories, five being too face--on for the velocity profile to be
well mapped, two having disturbed curves with strong distortion from a smooth
shape, and five (all of quality 2 or 3) either too poor in quality or too
sparse in coverage of the disk to be well fit.

Each calibrated spectrum was visually inspected and (rarely) single points
were removed from the fitted rotation curve if they were believed to be
contaminated by night sky lines or a cosmic ray residual.  A global
centerpoint was then measured for each galaxy, and the data were transformed
from wavelength to restframe velocity space.  This enabled the rotation curves
to be folded into a radial profile, so that they could be compared to the
luminosity profiles for agreement of structure and used to create mass models.
Velocity widths were measured to check for variations in the terminal
velocities relative to those measured from the 21~cm line, and to calculate
deviations from the Tully--Fisher relation.  Slopes were fit to the inner and
outer portions of each rotation curve and functional fits were also applied to
the form of the rotation curve, to permit quantitative comparisons between
galaxies.

The centerpoint of each spectrum in wavelength and on the sky was estimated in
two ways.  First, the wavelengths were binned into a histogram, and cuts were
made into each side at the 10\% level to remove scattered and outlying points.
The two endpoints of the trimmed histogram were averaged to determine the
cardinal central wavelength of the galaxy.  The spatial and wavelength
centerpoint of the rotation curve was then defined to lie at the location of
the nearest data point.  If the galaxy could not be traced through the nuclear
continuum in either \Halpha or \fnii (see Figure~\ref{fig:rcplt_smp}E), the
spatial centerpoint was allowed to ``float'' to a column with no data, and the
cardinal wavelength value was used as the wavelength centerpoint.  Second, the
point closest to the center--of--light (COL) determined from fitting the peak
of the galaxy continuum was chosen to be the spatial centerpoint, and the
wavelength of that point to be the central wavelength of the system.  

Default central values are those determined from the endpoints of the
histogram of wavelengths.  For 40 galaxies, however, the COL value is more
representative.  The spatial centerpoint for these galaxies differs by more
than two arcseconds when measured via the two techniques.  In the case of such
disagreement, the COL values were taken to be more accurate, and were used.
We also folded each rotation curve at the centerpoint to create two radial
profiles which were plotted on top of each other, and verified that the
curvature of the inner region of the rotation curve matched from profile to
profile.  In these cases, using the COL coordinates for centering decreased
the deviations between the two radial profiles.

The central wavelength was then converted to a restframe velocity, as
discussed in Vogt 1995.  The central redshift $z_c$ was taken to be the
redshift of the galaxy, corrected first to a heliocentric reference frame and
then to the reference frame of the Local Group according to the formulation

\begin{equation}
        cz_{LG} = cz_{\odot} + 300 \, \sin l \, \cos b
        \label{eq:loc_grp}
\end{equation}

\noindent
where $l$ and $b$ are the galactic longitude and latitude.  

\subsubsection{Velocity Widths} 

Velocity widths were measured in several ways from the optical spectra.
First, the velocity data were binned into a histogram, as above, and cuts were
made of a certain percentage into each side (the default being 10\%).  Though
this process bears an apparent resemblance to the fitting of a 21~cm profile
by making cuts in frequency space, it is only superficially analogous.  The
rotation curve is a spatially sampled curve, with no flux weighting given to
the data points.  A clipping algorithm is used to eliminate scattered outlying
points which deviate from the curve and high velocity points in the innermost
regions from bars (see Figure~\ref{fig:rcplt_smp}E); a curve which has leveled
off in the outer regions has a fairly constant velocity near the spatial
endpoints.  The velocity width of the galaxy was then taken to be the
difference between the velocity at each endpoint of the trimmed histogram.  A
second technique is designed for galaxies which are asymmetric or have
unbalanced arms (\eg strong \HII regions have been traced much further out
into the disk along one side than the other, and so a more extended velocity
profile can be derived, see Figure~\ref{fig:rcplt_smp}E).  The velocity data
were again binned, and cuts were made on each end.  Two velocity half widths
were defined by comparing the velocity of each endpoint with that of the
center of the galaxy as defined by the COL.  The full velocity width was then
set to be twice that of the half width measured further out along the disk.
In the case of \AGC210727/\CGCG097-125, for example, the velocity half width 
measured on the left side of the galaxy is 30 \kms higher than that measured 
on the right side, because it is based upon an inner peak in the velocity profile.  
We have thus chosen to double the half width measured from the right side of the
galaxy in determining the full width of the system.

\begin{figure} [htbp]
  \begin{center}\epsfig{file=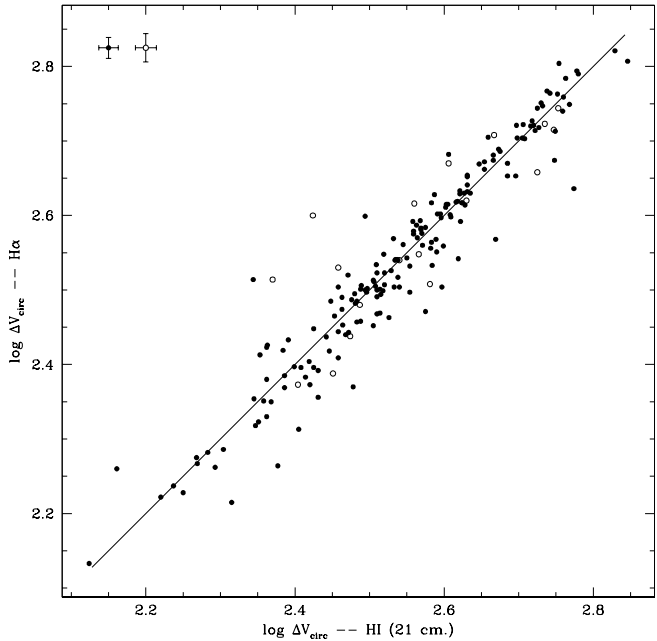,width=3.4truein}\end{center}
  \caption[Comparative Velocity Widths]
  {Velocity widths derived from both \Halpha optical spectra and 21~cm line
  profiles, showing a clear linear relationship.  Solid circles represent high
  quality data, and open circles data for which one or both of the velocity
  widths are derived from poor quality spectra.  The two points in the
  upper-left corner display representative median error bars for high and low
  quality data.  We find \dV(\Halpha) = (1.03 $\pm$ 0.02) $\times$ \dV({\rm
  H{\kern.2em{\smfont I}}}) $-$ (0.013 $\pm$ 0.008).}
  \label{fig:rc_rad_wd}
\end{figure}

A final velocity and velocity width were selected for each galaxy by examining
all of the data, taking into account the structure of the velocity profile.
The 12 curves of shape class 4, exhibiting solid body rotation, proved to be
better fit by measuring the endpoints of the total velocity histogram as the
velocity profile continued to rise appreciably in the outermost regions.  For
these galaxies the data were trimmed only of outlying deviant points, (\eg
isolated points at the end of the rotation curve with an error bar two to
three times higher than the norm lying more than 10 \kms above or below the
rest of the curve), and the histogram endpoints were then used to calculate
the velocity width.  For the other rotation curve shapes the standard trimmed
histogram technique functioned well.  It removed outlying points and a certain
amount of scatter, and minimized the effect of large velocities in the inner
regions upon the total rotation curve.  The key to any technique is a
consistent and uniform application, however, which has been achieved with the
current method.

It is interesting to compare velocity widths taken from optical spectra with
those determined from 21~cm line profiles (\cf Courteau's 1997 detailed
discussion of various techniques, and weighted comparisons, for extracting
widths from several samples in the literature).  We initially applied our
shape coding to the rotation curves within the subsample of Mathewson \etal
(1992) for which both optical spectra and 21~cm line profiles had been
processed (165 galaxies).  A comparison was made of the optical and radio
velocity widths, and a shape dependent correction factor applied to the
optical spectra on this basis for the well sampled types 1 through 3 (the
number of shape 4 curves is too small for a statistically meaningful
correction).  The correction is not very large, and as expected is largest for
the rotation curves which are rising in their outer regions: the optical
velocity width is reduced by 3 \kms for \QOS\ = 1, increased by 6 \kms for
\QOS\ = 2, and increased by 16 \kms for \QOS\ = 3.  Figure~\ref{fig:rc_rad_wd}
compares velocity widths measured via the two techniques for the current
sample, after this correction has been applied.  There is a good linear
relation over the full regime of interest, given the error budget in both
width estimates (see discussion in Giovanelli \etal 1997).  An average of
y-on-x and x-on-y weighted fits yields the relation (log \dV(\Halpha) - 2.2) =
(1.03 $\pm$ 0.02) $\times$ (log \dV({\rm H{\kern.2em{\smfont I}}}) - 2.2) $-$
(0.013 $\pm$ 0.008).  The scatter in the width difference relation is 31 \kms,
and remains fairly constant across the range of \dV (between 2.2 and 2.8),
comparable to previous results for comparable samples of spiral galaxies
(Courteau 1997).

One goal of our program is to explore changes in the fundamental parameters
(mass, size, luminosity) of galaxies as a function of environment.  We thus
need a robust measurement of the velocity width (mass) that is not affected by
\HI gas stripping (which can penetrate well into the central regions of the 
disk), nor by truncation of star formation (and dependent \Halpha emission) in
the outer regions of the disk.  We have thus also opted to measure the
velocity widths at a distance of two \Rd along the disks.  The rotation
curves were again folded about their centerpoints, and both the \Halpha and
\fnii data inspected.  These data were then interpolated to derive a
representative velocity width at two \Rd, after the disk scale lengths had
been corrected for the effects of galaxy inclination and for seeing
conditions.  

For 16 galaxies, the \Halpha flux is in absorption rather than in emission,
and the \fnii lines are too weak to be traced.  For these galaxies a small
correction was made to compensate for asymmetric drift, as the raw velocities
were measured from the stellar component.  Following the fitting technique
laid out in Neistein, Maoz, Rix, \& Tonry (1999), we fit the velocity width
and the dispersion of the \Halpha line at two \Rd, and derive a corrected
velocity,
\begin{equation}
  V_{cor}^2 = V^2(r) + \sigma^2(r) \, \left( 2 \frac{r}{R_d} - 1 \right)
  \label{eq:rix_fnc}
\end{equation}
where $r = 2$ R$_d$ (see also discussion in \pthree).  For a single galaxy,
\UGC8069, the velocity could also be measured from the \fnii flux;  the 
values agree to within 10 \kms.

\subsubsection{Rotation Curve Slopes} 

\begin{figure} [htbp]
  \begin{center}\epsfig{file=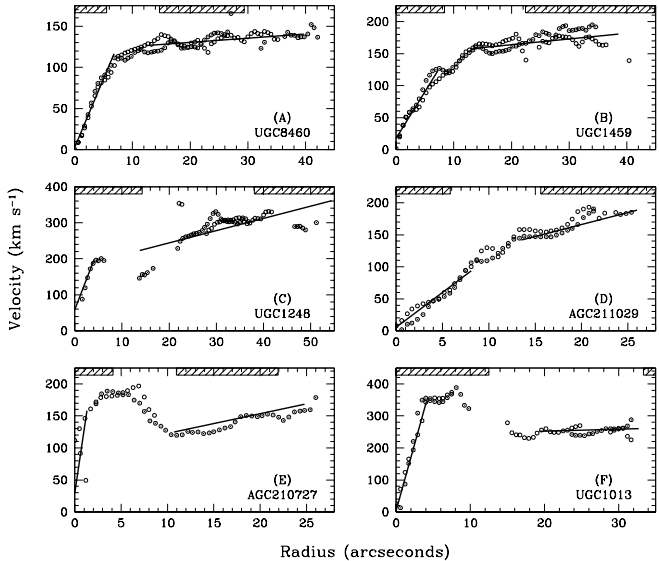,width=3.4truein}\end{center}
  \caption[Sample Rotation Curve Slopes]{Sample rotation curves of each shape
  class, overlaid with measured slopes (same galaxies as in
  Figure~\ref{fig:rcplt_smp}).  Rotation curves have been folded about the
  centerpoints; \Halpha circles are filled (approaching) and open (receding)
  to mark each side of the galaxy, and no error bars are plotted.  The slopes
  can be well quantified for galaxies $A$ -- $D$, (\QOS = 1 -- 4), but for
  those with inner peaks in the velocity profile the inner slope is heavily
  influenced by the peak behavior (often caused by a bar, as in $F$).
  Rectangular hatched boxes mark the zones 0 -- 15\% \RI\ and 40 -- 80\% \RI,
  over which the percentage rise of the curve is also measured.}
  \label{fig:rcplt_slps}
\end{figure}

Rotation curves can be defined to first order by a steeply rising linear inner
region, a short transition zone, and then a linear outer region which levels
off to a fairly constant level.  Shape differences arise from the relative
sizes of the regions, and when the outer region of a curve is truncated or
does not level off.  Each galaxy profile was first folded about its
centerpoint by taking the absolute value of positions and velocities relative
to that point.  Five points were marked on each rotation curve defining the
turnover point and a beginning and ending of an inner and an outer region,
each of approximately linear behavior (see Figure~\ref{fig:rcplt_slps}).  A
least--squares fit was made to each of the two regions, and these slopes were
normalized by the velocity half width, V(R$_f$), of the galaxy and the radius
of the turnover point, R$_{el}$.  This had the advantage of characterizing the
curves with parameters derived solely from the shape of the potential, rather
than utilizing an optical radius and adding in the effects of the light
distribution.  The turnover point is less strictly defined than an isophotal
radius, though, particularly for barred galaxies with additional substructure
in the inner velocity profile.  We define the slopes as
\begin{equation}
        S_{in} ~=~ \frac{\Delta V_{in} / V(R_f)}{\Delta R_{in} / R_{el}} 
        ~~~ \longrightarrow ~~~ 
        \frac{V(R_{el})}{V(R_f)}
        \label{eq:slp_in}
\end{equation}
and
\begin{equation}
        S_{out} ~=~ \frac{\Delta V_{out} / V(R_f)}{\Delta R_{out} / R_{el}} 
        ~~~ \longrightarrow ~~~  
        \frac{V(R_f) - V(R_{el})}{V(R_f)} ~\times~ \frac{R_{el}}{R_f - R_{el}}
        \label{eq:slp_out}
\end{equation}
and note their asymptotic forms for the simple case of two linear regions with
a narrow turnover zone (the inner slope reduces to the fractional rise of the
rotation curve within the turnover point).  In Figure~\ref{fig:rcplt_slps} we
overlay the slope measurements on sample rotation curves for all shape codes.
We observe that the inner and outer regions can be well defined and measured
for all of the galaxies, but that for the \QOS\ = 4 galaxy the turnover point
is difficult to constrain.  For the pictured two galaxies with inner peaks in
the velocity profiles, the inner slope is weighted heavily by that
substructure.

One can also normalize the ranges over which gradients are measured by
isophotal radii.  \Bb images, which have been used for this purpose in
previous studies (\cf Whitmore, Forbes, \& Rubin 1988, hereafter WFR; Amram
\etal 1993 \& 1996), are strongly influenced by current star formation;  
we have opted for \Ib images to be more sensitive to the older, underlying
disk population.  The cardinal \Bb \RB, the isophotal limit of 25 magnitudes
per square arcminute, lies roughly an order of magnitude above background
levels.  We compare \RB (taken from the RC3 for the 181 galaxies in common
with our sample) to \Ib isophotal radii and find a comparable limit at 23.5
magnitudes per square arcminute, \RI with a median offset of 0\arcsec.2
between the two radii (Vogt 1995).  We then use \RI to normalize the ranges
over which we fit velocity gradients to the rotation curves, for comparison
with the literature.  It was necessary to determine a rotational velocity for
a arbitrary points along each rotation curve.  Each rotation curve was folded
about its centerpoint, and the \Halpha and \fnii points from both sides of the
disk were binned together in a weighted average to produce a non-degenerate
radial curve.  These data were then smoothed with a low--pass FFT filter, and
velocities were calculated at specific points along the curve by using an
interpolating routine.  Inner and outer slopes of the rotation curves, which
we shall label as gradients to distinguish them from slopes as defined above,
were then defined to be the percentage rise of a synthetic, smooth rotation
curve fit to the data, in the ranges $0 - 0.15$ \RI and $0.4 - 0.8$ \RI.

Figure~\ref{fig:rcplt_slps} shows these fitting regions for a set of rotation
curves of various shapes.  As observed by WFR, galaxies which are
intrinsically brighter and/or of earlier type tend to rise more quickly to
their maximum velocity, with a short transition zone, and level off to a
constant velocity.  The 0.15 \RI point targets the inner region for most of
the galaxies, but (like the inner slope measurement described above) is
sensitive to inner peaks of the velocity profile.  The region over which the
outer gradient is measured can frequently extend beyond the terminal point of
the rotation curve, and thus when examining the outer gradients the sample
must be limited to those galaxies for which \RCf $\ge$ 0.8 \RI (roughly 50\%).
For barred galaxies such as \UGC1013 (F), the outer gradient range may extend
far beyond the extent of the rotation curve.

We can compare the distribution of $I$ based gradients within our sample with
the trends seen within previous data in \Bb to evaluate global trends, as long
as we do not assume a one--to--one correspondence between \RB and \RI for
individual sources.  There are no systematic offsets or trends between WFR,
Amram \etal and ourselves when common galaxies are compared; the outer
gradients agree to within the observational errors.  A primary finding of WFR,
that galaxies with falling rotation curves are found preferentially in cluster
cores, has not been reproduced.  Studies of optical and \HI velocity profiles
for galaxies in the Virgo cluster (Guhathakurta \etal 1988; Distefano \etal
1990; Sperandio \etal 1995), have not confirmed the trend, and 2D
Fabry--P\'{e}rot observations (Amram \etal 1993, Amram \etal 1996) of several
galaxies within the WFR sample did not find the same velocity gradients.  The
ranges over which each measurement of the velocity gradients was made were not
always the same, and as Amram notes, the inclusion or exclusion of single data
points along a moderate resolution rotation curve can have a significant
effect upon the measured gradient when there are few points.

\begin{figure} [htbp]
  \begin{center}\epsfig{file=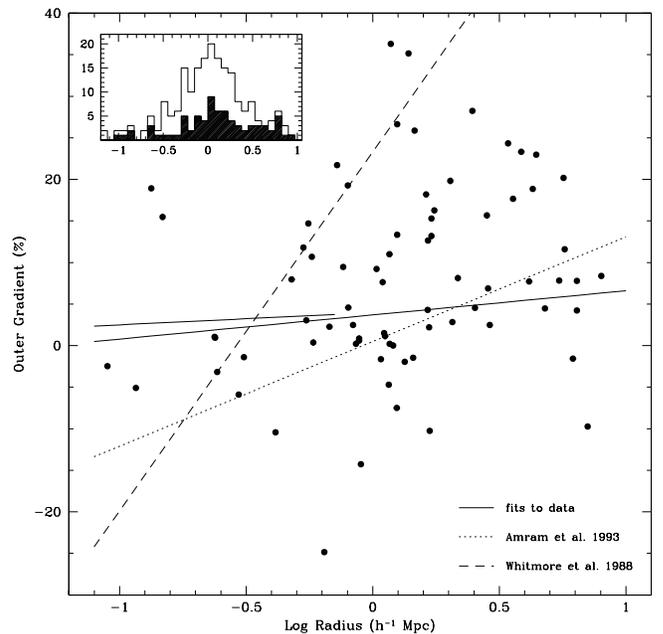,width=3.4truein}\end{center}
  \caption[Outer Gradient Trends]
  {The percentage rise in the \Halpha rotation curves between 0.40 and 0.80
  R$_{23.5}$ versus clustercentric radius.  The observed trend (solid lines)
  is extremely weak, whether fit to all radii or within 750 h$^{-1}$ Mpc, and
  we do not reproduce the stronger trends observed in smaller samples
  previously (dashed lines).  The inset histogram shows the distribution of
  galaxies within the sample as a function of clustercentric radius, with the
  fraction for which the rotation curve extends to 0.80 R$_{23.5}$
  highlighted.}
        \label{fig:rc_og}
\end{figure}

Figure~\ref{fig:rc_og} shows a plot of outer gradient as a function of cluster
radius, for 75 galaxies within 10 \hMpc of a cluster (foreground and
background galaxies have been suppressed).  The observed trend is exceedingly
weak, whether fit to galaxy gradients at all radii (OG = (2.9 $\pm$ 2.8)
$\times$ log R + (3.7 $\pm$ 0.4), correlation coefficient $r$ = 0.24), or to
those within 750 \hMpc from a cluster core (OG = (1.5 $\pm$ 8.8) $\times$ log
R + (3.9 $\pm$ 4.9), $r$ = 0.02), where one would expect environmental effects
to dominate.  In contrast, fits to samples of $\sim$20 galaxies have produced
OG = 43.2 $\times$ log R + 23.3 (WFR), and the far weaker trend of OG = 12.6
$\times$ log R + 0.5 (Amram).  Fits to our data produce an even shallower
slope than that found by Amram, and are fully consistent (within 1 $\sigma$)
with no correlation.  A re-analysis (Adami \etal 1999) of the data of Amram
found the observed trend to be stronger in late type spirals and not
significant in early types; we do not see the correlation in either type
group.

Rubin \etal argued that their evidence for declining rotation curves within
clusters implied that the affected sources had deficient halos, which had
either been stripped through an interaction with the cluster medium (\eg tidal
stripping, ram pressure sweeping) or with close neighbors (\eg harassment,
mergers) or had never been allowed to form.  This sample shows no evidence for
a strong correlation between outer gradient and clustercentric radius, and
thus provides no support for strong environmental variations within dark
matter distributions of cluster spirals relative to the field based on these
indicators.  We note that for all studies, the analysis has been restricted to
those galaxies for which \Halpha flux extends to a minimum of 0.80 \RI; the
accompanying histogram in Figure~\ref{fig:rc_og} shows that this removes a
significant fraction of observed galaxies at all radii.  The fraction rises to
60\% between 300 \hkpc and 3 \hMpc but is lower both in peripheral regions and
in cluster cores, suggesting that underlying sample bias (\eg what portion of
the total galaxy population falls within the complete sample) is also
relevant.  This factor could alter the observed effect (\eg stripped galaxies
have less extended \Halpha flux; see discussion in \ptwo).

\subsubsection{Rotation Curve Fits} 

\begin{figure} [htbp]
  \begin{center}\epsfig{file=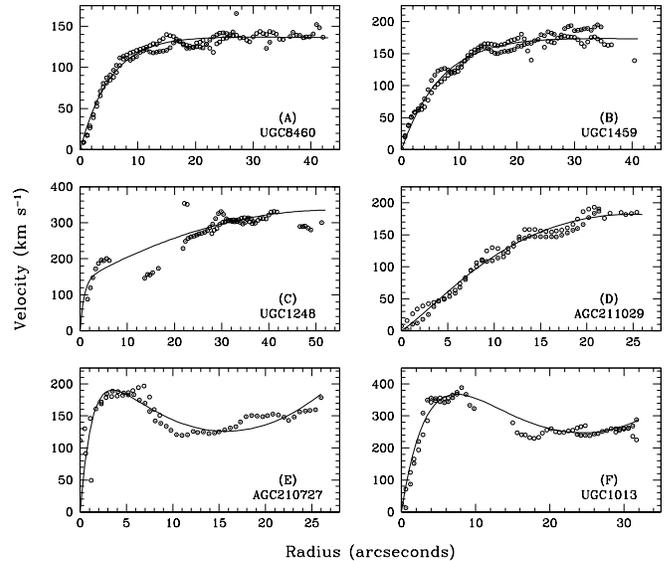,width=3.4truein}\end{center}
  \caption[Sample Rotation Curve Fits]
  {Sample rotation curves of each shape class, overlaid with measured fits
  (same galaxies as in Figure~\ref{fig:rcplt_smp}).  Rotation curves have been
  folded about the centerpoints; \Halpha circles and \fnii triangles are
  filled (approaching) and open (receding) to mark each side of the galaxy,
  and no error bars are plotted.  Galaxies of \QOS\ = 1 -- 4 can be well
  traced by the simple exponential form for the fit, as shown in $A$ -- $D$.
  Curves with a strong inner peak in the velocity profile require the higher
  order terms, and are more difficult to trace, as illustrated in $E$ -- $F$.}%
  \label{fig:rcplt_fits}
\end{figure}

Analytic functional fits were applied to the data to enable the fitting of
mass--to--light ratios.  A large scale fit, demonstrated in
Figure~\ref{fig:rcplt_fits}, was made to each rotation curve, of the form 
\begin{equation}
        V(r) = a_1 (1 - e^{a_2 r}) (1 + a_3 r  + a_4 r^2)
        \label{eq:rc_lrg}
\end{equation}
where the global form of the profile was determined by $a_1$ and $a_2$ (\ie a
two parameter fit), and $a_3$ and $a_4$ mapped the curvature in the outer
regions and could be set to zero to reduce the final term from quadrature to
linear or constant order.  The coefficients interact to create a model curve
such that $a_1$ scales with the maximum width of the curve, $a_2$ softens the
curvature of the turnover point and shifts it along the spatial axis, and
$a_3$ and $a_4$ control secondary curvature displayed primarily in the outer
regions of the curve.  Note that $a_1$ is {\it not} equivalent to the
asymptotic velocity of the profile, but can assume values much larger when a
rotation curve with a very soft turnover region is fit.  The fit can be
modeled on the total spectral data or on the \Halpha data alone; typically the
\fnii data were included unless they were of poor quality.  The purpose of the
analytic fit was to provide a quantitative smoothed form of the rotation
curve, so that a representative velocity could be calculated for an arbitrary
point along the disk.  The exponential form was chosen purely for convenience,
not because of a particular embodiment of physical variables, and is but one
of a number of possible and appropriate fitting models (\cf the Universal
Rotation Curve model of Persic \& Salucci 1990, model fitting by Rix \etal
1997, the arctangent function of Courteau 1997, Polyex fit of Giovanelli \&
Haynes 2002).

\subsubsection{Mass Distributions} 

The mass distribution of a galaxy can be determined solely from the observed
rotation curve by way of several simplifying assumptions (\cf Rubin \etal
1985).  By assuming that the observed rotation curves are a valid measurement
of circular, co-planar motion within the gravitation potential of the galaxy,
and that the mass distribution is spherically symmetric, the integral mass
distribution can be defined as
\begin{equation}
        M (r) = 2.3265 \times 10^5 \; V^2 (r) \; r \;\; M_{\odot}
        \label{eq:mm1}
\end{equation}
where $V$ is measured in \kms and $r$ in kpc.  Mass curves can then be defined
by taking the logarithm of the galaxy mass and radius, subtracting off the
offsets $M_m$ and $R_m$ from the centers of the curves, and creating a set of
curves centered about a single point.  Burstein \& Rubin (1985, see also
Burstein \etal 1986) determined for the sample of Rubin \etal such integral
mass curves, as a means of evaluating the total mass as a function of radius.
They defined three mass types of systematically increasing curvature.  The
type I mass curve was originally determined by fitting a sample of 21 field Sc
galaxies (Burstein \etal 1982), type II was created as an intermediate form
between I and III, and the mass curve for type III is that of a massless disk
in co-planar rotation with an isothermal sphere (Burstein \& Rubin 1985).
These three curves map out a fairly narrow range in the two-dimensional
$\left< M(v,r), r \right>$ space; it is interesting to note that almost all
spiral galaxies examined to date can be fit fairly well with one of these
three curves.  Note that as the derived mass depends linearly on radius, a
rotation curve which is flat at large radius (i.e. displays constant velocity)
will be transformed into a mass curve of linear slope in the outer regions.
Mass types were assigned for our sample of 329 galaxies by evaluating the
\Halpha rotation curves, the \fnii rotation curves in the cases where the
\Halpha data was deficient (due to weak tracing or strong absorption), and
model fits to the combined data sets.  It was found that there existed a small
number of galaxies with less curvature than the mass types I through III, and
thus a type 0 was created to hold these nine outliers.

\section{HI Observations} 
\label{sec:rd_obs}

Thus paper incorporates \HI data obtained in the course of this and other
projects, available in a digital archive.
The bulk of the \HI observations were made with the Arecibo Observatory
305--meter telescope.  The three northernmost clusters A426, A2197, and A2199
in the range $(39^{\circ} \le \delta \le 41^{\circ})$ have been observed
with the former 300--foot NRAO telescope in Green Bank (Freudling, Haynes,
\& Giovanelli 1988, 1992; Haynes, Magri \& Giovanelli 1988).  All galaxies
targeted for optical rotation curve studies were included in a dedicated 21~cm
line program (Haynes \etal 1997); the techniques of data acquisition and
analysis are discussed in detail therein and in Giovanelli \etal 1997.  A
total of 304 of the 329 galaxies were successfully observed, and 260 were
detected.  The data were binned according to quality, Q$_{HI}$.  The 245
quality (1) line profiles are well defined and well sampled, the 13 (2)
profiles are marginal but still usable for measurements of velocity width, the
two (3) spectra have detected fluxes but valid velocity widths cannot be
measured, the 44 (4) spectra are nondetections, and we do not presently have
good data on the 25 (0) objects (eight of which were observed but for which
the data are unusable due to strong interference or known source confusion).

The 21~cm line fluxes were calculated by correcting the observed flux,
measured at a level of 50\% of the profile horns, for the effects of random
pointing errors, source extent, and internal \HI absorption (Haynes \&
Giovanelli 1984; Giovanelli \etal 1995).  The \HI gas masses were calculated
by placing galaxy cluster members at the true cluster distance, and
non-members at individually derived redshift-dependent distances (see \ptwo\
for discussion of cluster membership criteria).  \HI deficiency is determined
by comparing the observed \HI gas mass with the expected \HI gas mass (Solanes
\etal 1996).  The expected \HI gas mass is calculated from galaxy morphological type
and linear diameter, using blue major axis diameters taken from the UGC (or
converted to UGC form from RC3 data) when available and otherwise estimated
from the Palomar Optical Sky Survey (POSS) prints.  Morphological typing was
done through visual examination of plate reproductions of the POSS-I prints at
Cornell University, which have a more extended dynamic range than do the
standard POSS prints, comparison with the UGC published type codes for UGC
galaxies, and (infrequently) updated upon detailed examination of our own \Ib
data.  The accuracy is of order $\pm\frac{1}{2}$ Hubble type overall.

For cases where no 21~cm line flux was detected, an upper limit has been
placed upon the source strength based on the rms noise per frequency channel,
$\sigma$, and the full velocity width, $\Delta V$, based on the optical
rotation curve of the galaxy.  We assume a simple linear flux model at a level
of 1.5$\sigma$ across the full velocity profile, such that
\begin{equation}
        S (max) = 1.5 \; \sigma \; \times \; \Delta V \;\; \mbox {Jy km s}^{-1}.
        \label{eq:HIdef3}
\end{equation}

\begin{figure} [htbp]
  \begin{center}\epsfig{file=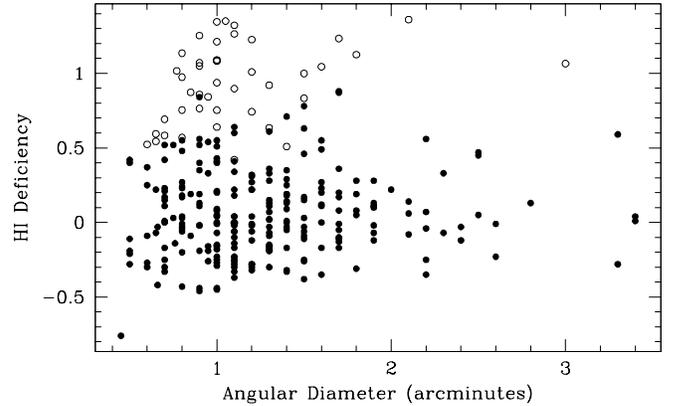,width=3.40truein}\end{center}
  \caption[HI Deficiency vs. Angular Diameter]
  {Plot of HI deficiency versus blue light angular diameters; there is no
  strong correlation.  Nondetections (observational lower limits on HI
  deficiency) are drawn as open circles; they have been redistributed using 
  survival analysis.}
  \label{fig:def_adiam}
\end{figure}

The application of this \HI deficiency estimation is problematic, given that
the calibrating field galaxies in the Solanes \etal sample have a larger
average angular diameter than those within this sample.  This incompleteness
will lead to an underestimation of the \HI deficiency, particularly for the
smallest (diameter $D_L \le 60''$) galaxies within the sample (see Giovanelli,
Chincarini, \& Haynes 1981).  Figure~\ref{fig:def_adiam}, however, confirms
that there is no strong trend in \HI deficiency with angular diameter.

A survival analysis was conducted on the distribution of nondetections, in
order to integrate them statistically into the detected sample.  The longterm
nature of our observing program leads to a substantial bias in the
distribution of nondetected sources in \HI deficiency, and also in \HI gas
mass, and the rms noise level of the spectra, as nondetections at the standard
flux limit were immediately identified and then preferentially re-observed
throughout the program.  This makes the distribution incompatible with a
random censoring model (\cf Magri \etal 1988).
We instead redistribute each nondetection into the portion of the total \HI
deficiency distribution which lies above the corresponding lower limit,
factoring in the statistical weight of the distribution.  Given the extreme
bias of the nondetections, they will dominate the tail of the distribution.
This technique will thus partially correct but still lead to an underestimate
of the \HI deficiency for the most extreme nondetections (\ie those lower
limits which fall near to or beyond the limit beyond which no true detections
are found within the sample).

\section{Photometry Observations} 

Optical images were taken using the KPNO 0.9--meter telescope and the MDM
1.3--meter telescope to obtain Mould \Ib data for the optical rotation curve
sample.  Full details of the data acquisition and analysis techniques are
provided elsewhere (Haynes \etal 1999, Giovanelli \etal 1997), as these data
represent only a subset of our large scale imaging program of nearby field and
cluster galaxies.  A total of 261 of the 329 galaxies were observed under
photometric conditions, and all but two of the images were deemed usable.  The
data were binned according to quality.  The 245 quality (1) images are have a
high S/N level and are well sampled, the six (2) images are of fair quality,
the two (3) images are marred by poor flat fielding, extensive bad columns, or
have too low a S/N level, and we do not presently have good, photometric data
on the 73 (0) objects.

The data were analyzed to provide surface brightness and luminosity profiles,
and estimates of ellipticity and position angle as a function of radius.  A
set of ellipses defining radii of constant surface brightness was fit to each
galaxy, and allowed to vary in eccentricity and in angular orientation about
the center of the galaxy.  A total magnitude was determined, typically to
within 0.05 magnitudes, for each galaxy by combining the flux from the
successive ellipses and extrapolating analytically to infinity, and a disk
scale length \Rd determined by the slope of the disk region of the surface
brightness profile.  A representative ellipticity and position angle were
selected by evaluating the mean and extremum values within the region
identified as the disk by visual inspection.

Disk scale lengths were determined for a face-on orientation, to compensate
for the varying effects of extinction with radius, and a correction was made
for seeing conditions.  As discussed in Giovanelli \etal 1994, a linear fit
was made to the surface brightness profiles over the range 0.40~\RI\ -- \RI,
where the measured value of \RI was first corrected to face-on orientation
based on the observed blue axial ratios.
This technique compensates for both inclination and redshift-dependent shifts
in the range over which the disk scale length is fit, as at large radii (\eg
where \RI is measured) disks can be treated as transparent, while the inner
regions suffer significantly from (inclination angle-dependent) extinction.
Disk scale lengths were thus fit over this disk regime for all galaxies; note
that the restriction to $r > 0.40$ \RI restricted the fit to the
disk-dominated region beyond the bulge for all spiral types.
A simple correction was then made for the effect of the seeing disk; which
could blur and flatten the surface brightness profiles when significant in
size relative to the width of the total disk fitting regime.  We measured the
FWHM of the estimated seeing disk, $FW_D$, for each exposure (determined from
the resolution of stellar images, and taken to be 1.5\arcsec if no calibration
data was available), and set
\begin{equation}
        R_d = \left( 1 - \frac{FW_D}{R_{23.5}} \right) \, R_{d} (obs).
        \label{eq:dsl2}
\end{equation}
An inclination angle has also been derived for each galaxy from the observed
ellipticity by convolving it with the estimated seeing disk, consistent with
the analysis outlined in Giovanelli \etal 1994.  

The galaxy luminosity has been transformed from an apparent to an absolute
magnitude by correcting for galactic extinction, applying a k--correction to
compensate for the redshift of the flux of the spectral band, correcting for
internal extinction, and then assuming a distance to the galaxy.  
Galactic extinction in the \Ib has been taken to be $A_I = 0.45 A_B$,
from the RC3 when available for the specific galaxy or approximated by that of
its nearest listed neighbors when necessary.  This correction is most important
for galaxies within the clusters A426 ($\overline{A_B} = 0.81$) and A539
($\overline{A_B} = 0.59$), which lie closest to the galactic plane; for the
rest of the sample $\overline{A_B} = 0.11$.  The fairly small k--correction has
been applied in the fashion of Han (1992), and the 
        \label{eq:magcor3}
Note that later studies (\cf Giovanelli \etal 1995) indicate that the form of
the correction for internal extinction is probably more complicated than this,
being dependent upon both the morphological type (earlier type spirals
suffering from more extinction than later type spirals) and the luminosity
class of the source.

\section{Mass Models}

Stellar mass-to-light ratios were estimated by fitting bulge, disk and halo
components to surface brightness and velocity profiles.  The \Ib surface
brightness profiles were first deconvolved into contributions from bulges and
disks, assuming two independent, discrete ellipsoidal components.  A series of
mass models, constraining light and dark matter in varied ways, were then fit
to combined \Halpha and \fnii\ major axis rotation curves; the ``maximum
disk'' model in particular was found to define the mass-to-light ratios in a
uniform and representative fashion, to within 10\%.  Significant trends were
found to be a function of expected variations with morphological type and
intrinsic luminosity.  When these effects were removed, the mass-to-light
ratios are relatively uniform.  The critical limitation proved to be the
spatial extent of the optical rotation curves.  A typical spiral galaxy in the
vicinity of a cluster does not have \Halpha flux extended sufficiently far out
along the disk to provide strong constraints on the disk and halo
contributions of the potential, unlike bright, nearby field galaxies.

\subsection{Bulge and Disk Decompositions} 

The \Ib surface brightness profiles were deconvolved into bulge and disk
components in a manner similar to that used by Kent (1986).  This avoids
parameterizing the bulge and disk components by way of set density laws (\eg
$r^{\frac{1}{4}} +$ exponential), allowing more freedom to conform to the
physical shape the profile.  The bulge and disk are instead modeled as
ellipsoids, each of constant axial ratio.  Major and minor axes profiles of
the galaxy are taken to be the sum of the two luminosity profiles, and given
that two components have different axial ratios the individual profiles can be
uniquely recovered.  We define $\mu(r)$ as the surface density profile and
$\epsilon(r)$ as the ellipticity, where
\begin{equation} 
	\epsilon(r) = 1 - \frac{b(r)}{a(r)}.
	\label{eq:epsdef}
\end{equation} 
The major and minor axis profiles are defined in terms of the bulge and disk
major axis profiles and ellipticities as 
\begin{equation}
	\mu_{maj}(r) = \mu_b(r) + \mu_d(r)
	\label{eq:majbd}
\end{equation}
\begin{equation}
	\mu_{min}(r) = \mu_b\left[\frac{r}{(1 - \epsilon_b)}\right] + 
		       \mu_d\left[\frac{r}{(1 - \epsilon_d)}\right]
	\label{eq:minbd}
\end{equation}
\noindent and can be combined to define the bulge profile as 
\begin{equation}
	\mu_b(r) = \mu_b^o[r \, (1 - \epsilon_b)] + \mu_b\left[r \, 
	\frac{(1 - \epsilon_b)}{(1 - \epsilon_d)}\right]
	\label{eq:majb}
\end{equation}
\noindent where $\mu_b^o(r)$ is the first order approximation of the bulge
profile of the form 
\begin{equation}
	\mu_b^o(r) = \mu_{min}(r) - \mu_{maj}\left[\frac{r}{(1 - \epsilon_d)}\right].
	\label{eq:majb0}
\end{equation} 
\noindent Equation~\ref{eq:majb} can be solved by calculating $\mu_b(r)$ and
introducing it iteratively into the righthand side of the equation to determine
the bulge profile from the total surface brightness profile along the major
axis;  convergence is typically achieved after only a few iterations.  
The sample extends to redshift $z = 0.045$, where 1\arcsec on the sky
corresponds to 0.6 \hkpc.  Each galaxy has been fit by a series of ellipses of
constant surface brightness at successive radii; due to constraints such as
the extreme rise in brightness in the galaxy core and the size of the seeing
disk, the innermost ellipse was typically placed at a radius of 1\arcsec .2 -
1\arcsec .4.  The derived profiles thus had to be extrapolated inwards from
this point.  Though the surface density profile is undetermined in this core
region, the total flux is known and constrains its form.

\begin{figure} [htbp]
  \begin{center}\epsfig{file=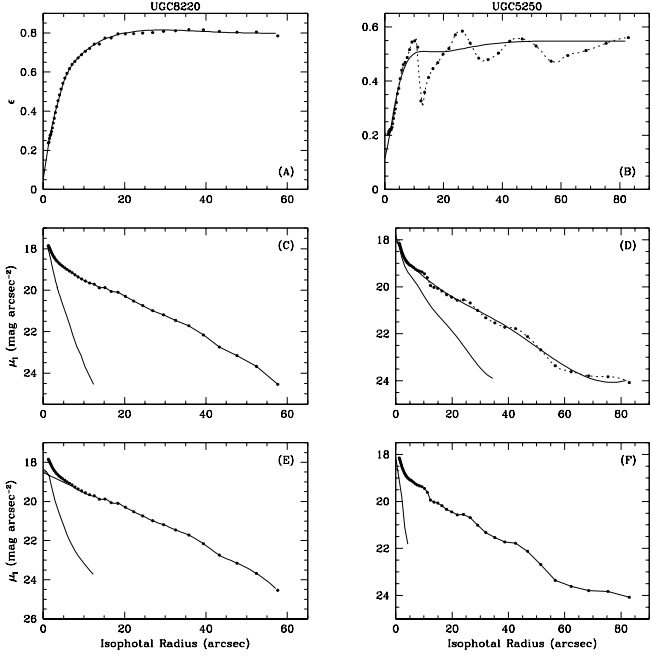,width=3.4truein}\end{center}
  \caption[Image Profiles]
  {Ellipticity profile and analytical fit (a) for \UGC8220 ($\epsilon_b =
  0.24$, $\epsilon_d = 0.83$), (b) fit by a smooth analytical function and by
  interpolation for \UGC5250 ($\epsilon_b = 0.21$, $\epsilon_d = 0.52$); 
  luminosity profiles divided into major and minor axis profiles (c) for
  \UGC8220, (d) fit by a smooth analytical function and by interpolation for
  \UGC5250 (interpolated from 1.9\arcsec outwards); deconvolved into
  bulge and disk components (e) for \UGC8220 (fit analytically in the inner
  6\arcsec.5, and by interpolation beyond), (f) for \UGC5250.}
  \label{fig:mm_buldsk}
\end{figure}

The ellipticity profile was first examined; profiles range from the radially
smooth to those with strong substructure, as shown in
Figure~\ref{fig:mm_buldsk}.  A fit was made to the profile of the form
\begin{equation}
	\epsilon(r) = a_1 + a_2 e^{-a_3 r} [1 + a_4 r + a_5 r^2 + a_6 r^3]
	\label{eq:eps1}
\end{equation}
\noindent selected by modeling the ellipticity profile upon the simple curve
\begin{eqnarray}
	\epsilon(r) & = & 1 - e^{-r}
	\label{eq:eps2}
\end{eqnarray}
\noindent with initial values for the fit parameters set to match the 
asymptotic behavior at the beginning and end of the profile.
The ellipticity of the bulge $\epsilon_b$ was taken to be the minimum
ellipticity in the central 5 \hkpc\ of the galaxy, and the disk ellipticity
$\epsilon_d$ to be the final ellipticity of the galaxy chosen by evaluating
the maximum, average, and representative user selected ellipticities.  In
certain cases the ellipticity profile varied considerably from this type of
smooth function; the profile was then modeled by an interpolating function
which estimated the profile at a given radius by utilizing the nearest five
points in the observed ellipticity profile).

The surface brightness profile along the major axis was fit in a similar 
fashion with the form 
\begin{eqnarray}
	\mu_{maj}(r) & = & a_1 + a_2 e^{-a_3 r} [1 + a_4 r + a_5  r^2 + a_6  r^3] \nonumber \\
	& + & a_7 e^{-a_8 r} [1 + a_9 r + a_{10} r^2 + a_{11} r^3]
	\label{eq:majsb1}
\end{eqnarray}
\noindent where the two exponential terms combine to fit both in the regime
near the galaxy center and in the asymptotic zone.  The resultant fit was
found to model the global behavior of the profile quite well, but failed to
reflect the small scale variations in the profile.  An interpolating function
was thus used to model all but the innermost parts of the profiles.  As the
bulge mass distribution is assumed to be smooth, it was calculated using the
smooth function (equation~\ref{eq:majb}).  The disk luminosity, however, was
calculated using the interpolating function to model the small scale structure
for all but the innermost regions.

The minor axis surface brightness profile was derived from the major axis
profile and the ellipticity profile.  The brightness profile was taken to be
that measured along the major axis, and the radius taken to be
\begin{equation}
	r_{min} = r_{maj} [1 - \epsilon(r_{maj})]
	\label{eq:minsb}
\end{equation}
\noindent and the profile $\mu_{min}(r)$ was then fit to the form of
equation~\ref{eq:majsb1}.  The surface brightness profile was then deconvolved
into bulge and disk components, according to equations~\ref{eq:majb} and
~\ref{eq:majb0}.

Once the bulge profile converged satisfactorily it was examined and terminated
at the first evidence of nonphysical behavior, in the form of secondary maxima
or discontinuities in the first derivative of the function.  This behavior
occurred only in the outer regions of the galaxy where the profile was
dominated by the disk;  thus the bulge profile could safely be set to zero from
this point outward and the disk profile set to be equal to the remainder of the
surface brightness profile.  The core region of the galaxy was then examined,
as the fit could at times exhibit nonlinear behavior in the inner 1\arcsec .5. 
Such cases were refit by eye to appropriate asymptotic behavior in this range.  

\subsection{Velocity Profile Models} 
\label{sec:vel_prf}

The next step was to calculate the velocity profile of the galaxy from the
gravitational potential of the different components of the galaxy, including
now a halo.  We calculate the velocity for the galaxy bulge, disk, and halo
components, defining 
\begin{equation}
	v^2(r) = r \, \frac{\partial \phi (r)}{\partial r}
	\label{eq:v2}
\end{equation}
\noindent where
\begin{equation}
	\phi(r) = \phi_b(r) + \phi_d(r) + \phi_h(r) 
	\label{eq:gp}
\end{equation}
and the potential of each component can be examined separately.  The bulge
potential can be modeled assuming a spherically symmetric distribution yielding 
\begin{equation}
	\frac{\partial \phi_b (r)}{\partial r} = G \frac{m(r)}{r^2}
	\label{eq:gp_b}
\end{equation}
\noindent where
\begin{equation}
	\rho(r) = - \frac{1}{\pi} \int_r^{\infty} \frac{\mu_b (s) \,ds}
						       {\sqrt{s^2-r^2}} 
	\label{eq:p_b}
\end{equation}
\noindent and
\begin{equation}
	m(r) = \int_0^r 4 \pi s^2 \rho (s) \, ds
	\label{eq:m1_b}
\end{equation}
\noindent which takes the form 
\begin{eqnarray}
	m(r) & = & \int_0^r 2 \pi s \mu_b (s) \, ds \\
	& + & \int_r^{\infty} \left[4 \sin^{-1} \left(\frac{r}{s}\right) - 
	\frac{4 r}{\sqrt{s^2-r^2}}\right] s \mu_b (s) \, ds.  \nonumber 
	\label{eq:m2_b}
\end{eqnarray}

The disk potential can be modeled as an infinitely thin disk (Toomre 1963).  
Beginning with Poisson's equation in cylindrical coordinates, 
\begin{equation}
	\frac{1}{r} \frac{\partial}{\partial r} (r \frac{\partial \phi}
	{\partial r}) + \frac{\partial^2 \phi}{\partial z^2} = 4 \pi G \rho = 
	4 \pi G \mu_d(r) \delta(z)
	\label{eq:poisson}
\end{equation}
a Fourier--Bessel transformation is applied to deduce the form
\begin{equation}
	\frac{\partial \phi_d(r)}{\partial r} = 2 \pi G \int_0^{\infty} s \mu_d(s) 
	\, H(s,r) \, ds
	\label{eq:FB}
\end{equation}
\noindent where
\begin{equation}
	H(s,r) = \left\{ \begin{array}{cl}
	  {\it \frac{2}{\pi s r} [K(k) - \frac{E(k)}{1-k^2}]} 
		& \; k = r/s \;\; (r<s) \\
	  {\it \frac{2 E(k)}{\pi r^2 (1-k^2)}} 
		 & \; k = s/r \;\; (s<r).
	\end{array}
	\right.
	\label{eq:g2}
\end{equation}
The Green's function H(s,r) has a logarithmically singular point at $s = r$.
This can be integrated numerically without difficulty by subdividing the
integral in Equation~\ref{eq:FB} into three parts, a region chosen to be
symmetric about the singularity, and the regions before and after this zone.

The dark matter halo potential can be modeled in a variety of ways.  A 
density law has been chosen of the form
\begin{equation}
	\rho = \frac{\sigma^2}{2 \pi G (r^2 + a^2)}.
	\label{eq:p_h}
\end{equation}
By modeling the halo as a sphere of radial symmetry in a matter analogous to 
that used for the bulge the following velocity profile is deduced.
\begin{equation}
	\frac{\partial \phi_h(r)}{\partial r} = \frac{2 \sigma^2}{r}
	\left\{1 - \frac{\tan^{-1}(r/a)}{(r/a)}\right\}
	\label{eq:v2_h}
\end{equation}

\subsection{Luminosity Corrections}
\label{ext-disk} 

The effects of extinction upon both the luminosity profile and the optical
velocity profile are fairly complicated to model.  It has been shown
(Giovanelli \etal 1994; Giovanelli \etal 1995) that the effects of extinction
on a luminosity profile vary with total luminosity and morphological type; in
particular, early type spiral seem to suffer a greater extinction than later
types.  Extinction also varies as a function of galaxy radius, and thus can
significantly alter the slope of the luminosity profile.  The main effect upon
an optical velocity profile is that the inner regions of the galaxy tend to be
obscured by dust, particularly in the case of a highly inclined galaxy, and
the observed velocity profile may not be representative of true rotation.  The
inner slope is dependent upon galaxy inclination, due to extinction effects
(Giovanelli \& Haynes 2002).  This does not impact the mass modeling
technique used here significantly, however, as the model parameters are most
strongly controlled by the behavior of the rotation curve at large radii (\ie
the amplitude of the profile beyond the turnover point).

A correction has been applied to the luminosity profile to account for the
projection effects of viewing the disk luminosity from an inclined position.
In the case of a sphere, the observed luminosity profile is representative of
the true radial luminosity profile, but in the case of an optically thin disk
the apparent profile flux is increased by a factor of $(1 - \epsilon)$, the
minor to the major axis ratio, over the true profile.  This effect was dealt
with by considering the two extremum cases.  In the first case the galaxy was
treated as a pure thin disk system, and all flux was corrected for projection.
In the second case a dividing point was set at the radius at which the galaxy
becomes a pure disk.  The flux beyond this point was corrected for projection,
while the inner flux was treated as if it came from a pure sphere and no
correction was applied.  The two cases were then averaged together to achieve
a full correction of the pure disk zone and a partial correction in the region
of combined bulge and disk, and this correction was applied to the data.

\subsection{Mass to Light Ratios} 

The mass-to-light ratios of the bulge and disk are defined as \mlb and \mld, in units of
$\left[\frac{M_{\odot}}{L_{\odot}}\right]$.  A set of models were chosen to
explore the parameter space defined by \mlb, \mld, and the halo parameters
$\sigma$ and $a$ (Vogt 1995).  Attempting to fit all four variables
simultaneously with no additional constraints results in a poorly constrained
search of a large parameter space which can deviate wildly from physically
reasonable parameters; thus we have selected a set of physically
representative models with varying constraints.

The first three models are pure disk and bulge models.  Model 1 fits \mlb and
\mld to the full velocity profile; the overall importance of the bulge and
disk can be evaluated by comparing their amplitudes to those found in models
which contain significant halo components.  Model 2 forces \mlb to zero and
fits \mld alone to the data and model 3 holds \mlb and \mld equal, giving
information on the dependencies between the bulge and the disk.

The remaining models all include a halo component.  Model 4 assumes a constant
density halo, which is physically reasonable if the optical rotation curves do
not extend out beyond the inner portion of the halo exhibiting solid-body
rotation. Our data do not support this assumption, but the model is useful for
comparison to literature (\eg Kent 1986), and has the simplifying advantage of
parameterizing the halo with a single variable.  Model 5 determines $\sigma$
($\sigma \equiv \frac{V}{\sqrt(2)}$) from the terminal velocity of the \HI
line profile.  This assumes that the velocity profile is relatively flat
beyond the optical radius, and is strongly dominated by the halo.  For those
galaxies in the sample where no \HI width was determined (because of \HI
deficiency or because the galaxy was not observed), the terminal velocity of
the optical rotation curve was used.  This model has been labeled informally
the ``maximum halo" or ``maximum dark" hypothesis because the obvious result
of setting $\sigma$ to reflect the terminal velocity of the profile is to
increase the halo, at the expense of the disk.
\label{max-disk}
Model 6 is commonly referred to as the ``maximum disk" hypothesis (van Albada
\etal 1985), though more accurately designated the ``maximum light".  An inner
region is defined within the peak of the disk velocity profile, and \mlb and
\mld are fit to the velocity data within this point.  Then $\sigma$ and $a$
are fit to the remnant over the entire profile.  This assumes that the mass
distribution of the inner part of the galaxy, well within the optical radius,
is dominated by the stellar components.  Model 6 has been chosen as the most
representative mass model for the sample.  Model 7 submerges the bulge
component of the galaxy into the disk.  The fit to the inner region (as
defined in model 6) is done for \mld alone.  This model is completely
independent of the bulge-disk deconvolution, and a measure of the importance
of the bulge component can be gained by comparing it to models 1 and 6.  Model
8 fits a pure halo to the entire velocity profile, an extreme model which
serves only to constrain the range of possible halo contributions.

\begin{figure} [htbp]
  \begin{center}\epsfig{file=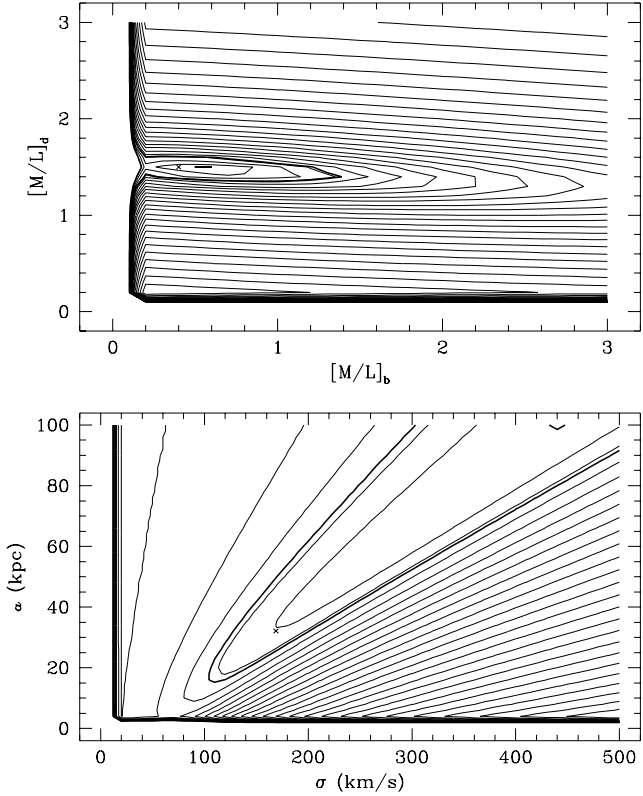,width=3.4truein}\end{center}
  \caption[$\chi^2$ values within $\frac{M}{L}$ space]
  {A mapping of $\chi^2$ in (a) $< \left[\frac{M}{L}\right]_d, 
  \left[\frac{M}{L}\right]_b >$ space and (b) $\left< \sigma, a \right>$ space
  for \UGC8220.  Contours are logarithmic, and heavy lines delineate a
  $\chi^2$ of one and of two.  The final values are \mld = 1.5 and \mlb = 0.4,
  $\sigma = 169$ km s$^{-1}$ and $a = 32$ kpc (marked with crosses on plots).
  $\left[\frac{M}{L}\right]_d$ is well defined, $\left[\frac{M}{L}\right]_b$
  less so, and there exists a complete family of halo parameters with
  equivalent validity, illustrating the difficulty in constraining the halo
  from these type of data}
  \label{fig:mm_chi2}
\end{figure}

Figure~\ref{fig:mm_chi2} shows the distribution of $\chi^2$ for fits of \mld,
\mlb, $\sigma$, and $a$ for \UGC8220 under the maximum disk model.  The 
spread in the mass-to-light ratios can be taken as a valid indicator of the sum
of the errors in the analysis and the intrinsic scatter.  The halo parameters
are not well constrained, however.  Due to the limited radial extent of the
rotation curves, there exists a family of solutions within the two-dimensional
$\left< \sigma, a \right>$ space which work equally well.  The chosen
solutions have been selected to produce terminal velocity behavior in
agreement with the terminal velocity determined from the \Halpha and \HI
spectra, and to avoid an extremely abrupt rise to maximum amplitude (very
small values of $a$).  While the former practice is linked to valid
observables, the second is an arbitrary modification designed to produce an
standardized technique which can be applied uniformly.  Recall, however, that
under the maximum disk hypothesis the mass-to-light ratios are set {\it before}
the halo parameters and are thus unaffected by them.  We emphasize that the
poor constraint upon halo parameters is an endemic problem for all mass models
determined from longslit \Halpha spectra, and not peculiar to our sample in
any fashion -- \NGC3198 is prominently featured in the literature precisely
because of its extended velocity profile, but is not representative of the
general population of spiral galaxies..

We can combine the models of the bulge and the disk profiles to define the
stellar mass-to-light ratio as
\begin{equation}
	\left[\frac{M}{L}\right]_{\star} = \frac{\mmlb L_b + \mmld L_d}{L_b + L_d}
	\label{eq:MLstar}
\end{equation}
\noindent
and incorporate the halo to derive the ratio of dark-to-light matter within 
the luminous portion of the galaxy as
\begin{eqnarray}
	R_M & = & \frac{M_h}{M_b + M_d}	\nonumber \\ [0.10in]
	    & = & \frac { 2 \sigma^2 r_2
		  \left\{1 - \frac{\tan^{-1}(r_2/a)}{(r_2/a)}\right\} }
		  {r_1 v_b^2 (r_1) + r_2 v_d^2 (r_2)}
	\label{eq:Mdl_rat}
\end{eqnarray}
\noindent
where r$_1$ is the radius of the final point of the bulge (where the bulge
luminosity profile terminates), v$_b$ the calculated velocity of the bulge
component, r$_2$ is the radius of the final point of the luminosity profile,
and v$_d$ the calculated velocity of the disk component.  This places a lower
limit upon the mass of the halo, as it assigns to the halo the mass necessary
to produce the observed rotation curve after calculating the components due to
the bulge and disk velocity profiles.

\subsubsection{Error Estimations}
\label{sec:errest}

The sources of error inherent in mass modeling can be divided into several
categories, namely the observed luminosity and velocity profiles, the derived
ellipticity and inclination angle, the deconvolution of the luminosity
profile, and the selection of the mass-to-light ratios and halo parameters to
best match the velocity profile.  This set of errors combines with the
variations with independent galaxy properties as yet uncorrelated to produce
the observed scatter.
The case of a galaxy with a rotation curve which ends significantly within the
optical radius is worthy of special notice.  If the rotation curve does not
reflect the true full amplitude of the velocity profile, an artificially low
mass and mass-to-light ratio will be derived.  A simple test for this problem
which has been performed for the sample is a comparison of the optical and
radio velocity widths for suspect galaxies, as the radio widths will not have
been similarly affected; no problem is revealed.

Errors in luminosity profile deconvolution are relatively unimportant.  They
become significant in the case of a galaxy where the bulge and disk
ellipticities are very similar and the profiles are thus difficult to separate
or when barlike structures in the inner regions of the galaxy create a
luminosity profile which cannot be easily separated into two components of
constant ellipticity.  For most galaxies a variation of the deconvolution has
an effect on the bulge mass-to-light ratio but virtually no effect on the disk
mass-to-light ratio, as it is determined from the entire profile and thus
minimally affected by the deconvolution occurring within the inner regions.

The error on \mld is on the order of 10\%, and in general can be constrained
quite well by the outer regions of the optical rotation curve.  The small
number of sources for which \mld is less than 0.3 are galaxies for which an
unhappy combination of noncircular velocity behavior in the inner regions
which is poorly matched by either the bulge or the disk and weak sampling in
the outer regions creates an invalid best fit solution.  As the inner regions
can be subverted far more easily by distortion due to extinction effects on
the luminosity profile {\it and} on the rotation curve, and the effect of
noncircular velocities, \mlb is less well defined.  A poor deconvolution will
also have a far more significant effect upon \mlb than on \mld, because the
bulge is present only in the overlap region.  Figure~\ref{fig:mm_chi2} maps
the value of $\chi^2$ in the two-dimensional $\left< \mmlb, \mmld \right>$ and
in space and in $\left< \sigma, a \right>$ space for \UGC8220.  It is clear
that \mld is constrained much more strongly than \mlb.  The shift in \mlb away
from the nadir of the map is caused by a slight mistracing of the inner slope
of the rotation curve by its fit function, easily evaluated by the human eye.
The halo parameters $\sigma$ and $a$ together define a family of solutions
which can match the velocity profile equally well within the optical rotation
curve radius but diverge widely from each other in their asymptotic behavior
at significantly larger radii.

\begin{figure} [htbp]
  \begin{center}\epsfig{file=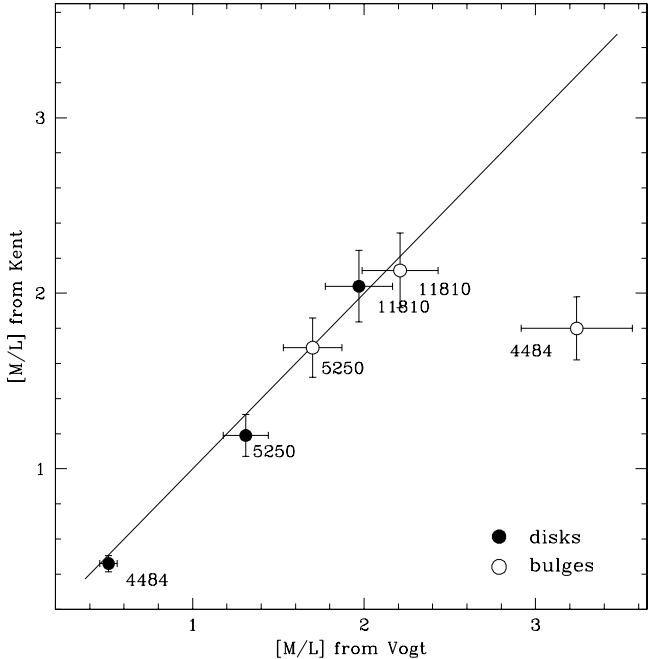,width=3.4truein}\end{center}
  \caption[Comparative M/L Ratios]
  {A comparison of [M/L] ratios within the overlap of the sample analyzed by
  Kent (1986).  Dark circles represent disks and open circles bulges.  The
  deviation of the bulge of \UGC4484 can be explained by different
  interpretations of the noncircular structure in the nuclear region and by
  the variation in the two luminosity profiles in the inner core within the
  seeing disk.  The line denotes $y = x$ and error bars are plotted for a 10\%
  error (representative of the total accuracy of the derived [M/L] ratio, not
  simply of the fitting technique).}
  \label{fig:mm_kent}
\end{figure}

Figure~\ref{fig:mm_kent} shows the good agreement between calculated
mass-to-light ratios for the overlap of this sample and that analyzed by Kent
(1986).  Kent's data has been corrected to the distances and inclination
angles used herein, the latter being a key factor, assuming that his
inclination angle was calculated from the tabulated values of the disk
eccentricity in his Table 3 and with an intrinsic axial ratio of 0.2.  The
data are presented in bolometric units of $\left[ \frac{M_{\odot}}{L_{\odot}}
\right]$ where no extinction correction has been applied (Kent's photometric
profiles are \Rb, while ours are I-band).  The deviation of the bulge of UGC
4484 can be explained by different interpretations of the noncircular
structure in the nuclear region and by the variation in the two luminosity
profiles in the inner core within the seeing disk (Kent quotes a seeing FWHM
of 2$''$.5, versus 1\arcsec .2 herein).

\section{Data Presentation}
\label{ch:data_all}

We present in Tables~2 through 4\footnote{ The complete version of these tables
is available in the electronic edition of the Journal.  The printed edition
contains only a sample of each, for guidance regarding their form and
content.} a listing of key galaxy properties for this sample of 329 local
northern hemisphere galaxies, sorted by Right Ascension.  Note that quantities
are not necessarily identical to similar parameters from previous
publications, though no significant differences exist.  A small fraction of
the photometric data for the Coma cluster have been taken by combining
literature results with our own measurements; specifics of those observations
can be found for those sources within Giovanelli \etal 1997).  Because of
length, specific comments on individual galaxies are placed separately from
the table, in the main text below.  The data are as follows.

{
\parindent 0.0in

\begin{table*} [hbtp]
  \caption{Galaxy Parameters I}
  \begin{center}
  \begin{tabular} {l r r c r l c r r r c r} 
  \tableline
  \tableline
  Names & \multicolumn{1}{c}{R.A.} & \multicolumn{1}{c}{Dec.} & T & \multicolumn{1}{c}{$V_{LG}$} & 
  \multicolumn{2}{c}{Cluster} & \multicolumn{1}{c}{$r$} & \multicolumn{1}{c}{$s$} & \multicolumn{1}{c}{h$^{-1}$ kpc} & Qual & 
  \multicolumn{1}{c}{$\theta$} \\
  & \multicolumn{1}{c}{h m s} & \multicolumn{1}{c}{d m s} & \multicolumn{2}{r}{{\small km s$^{-1}$}} & 
  & & \multicolumn{1}{c}{$^{\circ}$} & \multicolumn{1}{c}{$\sigma$} & \multicolumn{1}{c}{$\prime\prime$} & & 
  \multicolumn{1}{c}{$^{\circ}$} \\ 
  \multicolumn{1}{c}{(1)}  & \multicolumn{1}{c}{(2)}  & \multicolumn{1}{c}{(3)}  & \multicolumn{1}{c}{(4)}  & \multicolumn{1}{c}{(5)}  & 
  \multicolumn{1}{c}{(6)}  & \multicolumn{1}{c}{(7)}  & \multicolumn{1}{c}{(8)}  & \multicolumn{1}{c}{(9)}  & \multicolumn{1}{c}{(10)} & 
  \multicolumn{1}{c}{(11)} & \multicolumn{1}{c}{(12)} \\
  \tableline 
  U00809/520-033 & 01 13 00.0 & +33 33 00 & 5  &  3920 & N507   & f & 1.54 & -2.00 & 0.190 & 1311 &  23 \\
  U00841/502-025 & 01 16 22.3 & +32 46 06 & 4  &  5287 & N507   & c & 0.82 &  1.08 & 0.233 & 1211 &  54 \\
  U00927/N496    & 01 20 22.2 & +33 16 04 & 4  &  5722 & N507   & c & 0.22 &  2.06 & 0.233 & 1211 &  28 \\
  U00940/521-016 & 01 20 40.0 & +34 18 30 & 5  &  6707 & N507   & b & 1.25 &  4.28 & 0.325 & 1211 &  85 \\
  U00944/N512    & 01 21 10.5 & +33 38 54 & 2  &  4587 & N507   & c & 0.63 & -0.50 & 0.233 & 1111 & 116 \\
  \tableline \\
  \multispan{12}{The complete version of this table is in the electronic edition of the Journal.  The printed edition} \\
  \multispan{12}{contains only a sample.} \hidewidth
  \end{tabular}
  \end{center}
  \label{tab:crdnl1a}
\end{table*}

\vskip 0.15truein
{\bf Table~2} (Galaxy Parameters I)

{\bf Col. 1:} Source identification.  The first listing is an internal coding
number, which is equal to the \UGC number if the galaxy is listed in the \UGC
catalog (Nilson 1973).  A second galaxy name is added with the following
priority scheme: if the galaxy has an \NGC or \IC number, that coding is entered
preferentially; if the galaxy is listed in the \CGCG (Zwicky, Herzog, \& Wild
1968), that coding is listed with second priority, in the form: field number
-- ordinal number within the field.  If entries in those three catalogs are
not available, a second name is not entered.

{\bf Cols. 2 and 3:} Right Ascension and Declination in the 1950 epoch; 
the listed positions typically have 15\arcsec\ accuracy.

{\bf Col. 4:} Morphological type code in the RC3 scheme, where code 1
corresponds to Sa's, code 3 to Sb's, and code 5 to Sc's. The type code is
followed by a ``B'' when the galaxy disk has an identifiable bar.

{\bf Col. 5:} The galaxy radial velocity (redshift) in the CMB reference
frame; this value is obtained from high quality ($Q_{HI}=1$) \HI measurements
where available, and otherwise taken from the optical spectra.

{\bf Col. 6:} The cluster with which the galaxy is associated (left blank for
field sources).  

{\bf Col. 7:} Membership code.  In most cases the code consists of one of four
letters: {\it c, g, f, b,}. Code {\it c} signifies that the galaxy is
considered a {\it bona fide} cluster member. Code {\it g} means that the galaxy
is sufficiently removed from the cluster center that cluster membership cannot
be safely assigned; the galaxy is however in the periphery of the cluster, at
roughly the same redshift and it is a nearby supercluster member. Codes {\it f}
and {\it b} identify foreground and background galaxies respectively. The
criteria adopted to arrive at the membership assignment are discussed in
\ptwo. In two clusters, additional codes had to be used, due to the complex
nature of the structure.  The Cancer group has been recognized to be an
assembly of clumps, dubbed A, B, C, D (Bothun \etal 1983). We use the criteria
given in that reference to assign membership to each of the clumps.  The case
of A2634 ($V_{LG} \sim 9200$ \kms) and A2666 ($V_{LG}\sim 8100$ \kms) is also
complicated.  Scodeggio \etal (1995) have produced a detailed study of the
region, and within the redshift range spanned by the two clusters they also
distinguished several other groups. In particular, one of the groups is at a
systemic velocity near 7,000 \kms, near in redshift to, but distinct from
A2666. We reserve the code ``g7'' for the members of the 7,000 \kms group
projected on the foreground of A2634.

{\bf Cols. 8 and 9:} The angular distance $r$ in degrees and the radial
separation $s$ in $\sigma$, from the center of each cluster.  As the Cancer
group consists of several subclumps, distances have been measured from the
center of the main clump A (see Bothum \etal 1983).

{\bf Col. 10:} The angular size of 1 \hkpc, in arcseconds.

{\bf Col. 11:} Quality and shape codes.  A four digit code listing \QO, the
quality of the optical spectra, \QOS, the shape of the rotation curve, \QH,
the quality of the 21~cm line spectra, and \QI, the quality of the \Ib
photometry.

{\bf Col. 12:} The position angle $\theta$, measured eastwards from north,
along which the optical spectrum was observed, in degrees.

\vskip 0.15truein
{\bf Table~3} (Galaxy Parameters II)

{\bf Col. 1:} Source identification, exactly as in Table:~2.  

{\bf Cols. 2 and 3:} The turnover (elbow) point and the maximum radial
extent of the rotation curve \RCf, in arcseconds.

{\bf Cols. 4 and 5:} The differential extent of \Halpha or {\rm [N\II]}
emission, in \hkpc, followed by the ratio.  

{\bf Cols. 6 through 9:} The measured inner and outer slopes $S_{in}$ and
$S_{out}$ in normalized units, and inner and outer gradients $G_{in}$ and
$G_{out}$ as a percentage of the measured optical velocity half--width, of the
rotation curves (with estimated uncertainties between brackets).

 
{\bf Cols. 10 through 13:} The velocity width from the optical spectroscopy,
in \kms.  Col. 10 lists the measured velocity width, col. 11 the velocity
width corrected for the shape of the rotation curve and cosmological stretch,
and col. 12 the corrected velocity width converted to edge--on viewing (with
its estimated uncertainty, taking into account both measurement errors and
uncertainties arising from the corrections, between brackets).  Col. 13 lists
the corrected velocity width converted to edge--on viewing, measured at 2
R$_d$.

\begin{table*} [htbp]
  \caption{Galaxy Parameters II}
  \begin{center}
  \begin{tabular} {l r r r r r r r r c r r r r} 
  \tableline
  \tableline
  Names & \multicolumn{1}{c}{R$_{el}$} &  \multicolumn{1}{c}{R$_f$} & \multicolumn{1}{c}{$\Delta$H$\alpha$} &  \multicolumn{1}{c}{H$\alpha _{rat}$} & 
  \multicolumn{1}{c}{S$_{in}$} & \multicolumn{1}{c}{S$_{out}$} & \multicolumn{1}{c}{G$_{in}$} & \multicolumn{1}{c}{G$_{out}$} & 
  \multicolumn{1}{c}{OW$_0$} & \multicolumn{1}{c}{OW$_1$} & \multicolumn{1}{c}{OW$_2$} & \multicolumn{1}{c}{OW$_3$} \\
  & \multicolumn{1}{c}{$\prime\prime$} & \multicolumn{1}{c}{$\prime\prime$} & \multicolumn{2}{l}{h$^{-1}$ kpc} & & & & & 
  \multicolumn{1}{c}{--} & \multicolumn{2}{c}{km s$^{-1}$}  & \multicolumn{1}{c}{--} \\ 
  \multicolumn{1}{c}{(1)}  & \multicolumn{1}{c}{(2)}  & \multicolumn{1}{c}{(3)}  & \multicolumn{1}{c}{(4)}  & \multicolumn{1}{c}{(5)}  & 
  \multicolumn{1}{c}{(6)}  & \multicolumn{1}{c}{(7)}  & \multicolumn{1}{c}{(8)}  & \multicolumn{1}{c}{(9)}  & 
  \multicolumn{1}{c}{(10)} & \multicolumn{1}{c}{(11)} & \multicolumn{1}{c}{(12)} & \multicolumn{1}{c}{(13)} \\ 
  \tableline 
  U00809/520-033 & 10.9 & 39.2 &  1.43 &  0.81 & 0.66(04) &  0.10(02) &  55 &   13 & 270 & 282 & 286( 6) & 257( 7)  \\
  U00841/502-025 & 13.3 & 31.7 &  2.14 &  0.71 & 1.14(05) &  0.03(06) &  73 &   -8 & 247 & 245 & 249(10) & 254(12)  \\
  U00927/N496    &  6.8 & 39.8 &  2.28 &  0.75 & 0.96(06) &  0.04(01) &  70 &   21 & 270 & 268 & 318(13) & 318(14)  \\
  U00940/521-016 &  8.8 & 30.6 &  0.94 &  0.91 & 0.97(04) &  0.01(01) &  48 &    0 & 282 & 278 & 352( 9) & 317( 8)  \\
  U00944/N512    &  7.7 & 41.0 &  0.96 &  0.90 & 0.83(05) & -0.01(01) &  65 &  -10 & 515 & 500 & 506( 5) & 510( 6)  \\
  \tableline \\
  \multispan{13}{The complete version of this table is in the electronic edition of the Journal.  The printed edition contains } \\
  \multispan{13}{only a sample.} \hidewidth
  \end{tabular}
  \end{center}
  \label{tab:crdnl1b}
\end{table*}

\vskip 0.15truein
{\bf Table~4} (Galaxy Parameters III)

{\bf Col. 1:} Source identification, exactly as in Table:~2.  

{\bf Cols. 2 through 4:} The velocity width from the 21~cm line profiles, in
\kms.  Col. 2 lists the measured velocity width, measured at a level of 50\%
of the profile horns, col. 3 the velocity width corrected for instrumental and
data processing broadening, signal to noise effects, interstellar medium
turbulence, and cosmological stretch, and col. 4 the corrected velocity width
converted to edge--on viewing (with its estimated uncertainty, taking into
account both measurement errors and uncertainties arising from the
corrections, between brackets).

{\bf Cols. 5 through 6:} The observed and the predicted \HI masses, in units
of $\log 10^{10} M_{\odot}$.

{\bf Col 7:} The adopted inclination $i$ of the plane of the disk to the line
of sight, in degrees, (90$^\circ$ corresponding to edge--on perspective).

{\bf Col 8:} The apparent magnitude, to which k--term, galactic and internal
extinction corrections were applied, measured in Mould \I.

{\bf Col 9:} The absolute \I magnitude, computed assuming that the galaxy is
at the distance indicated either by the cluster redshift, if the galaxy is a
true cluster member, or by the galaxy redshift if it is not. The calculation
assumes $H_\circ = 100h$ \kms Mpc$^{-1}$, so the value listed in column 16 is
strictly $M_I + 5\log h$. In calculating this parameter, radial velocities are
expressed in the LG frame and uncorrected for any cluster peculiar motion.
The uncertainty on the magnitude, indicated between brackets in hundredths of
a mag, is the sum in quadrature of the measurement errors and the estimate of
the uncertainty in the corrections applied to the measured parameter. The
error estimate does not include the uncertainty on the value of the distance.

{\bf Col 10:} The disk percentage of the total luminosity.

{\bf Col 11:} The disk exponential scale length, \Rd, corrected to face--on
value, in arcseconds.

{\bf Col 12:} The radius \RI\ at which the surface brightness profile reaches
a limiting value of 23.5 magnitudes per square arcsecond, in arcseconds.

{\bf Cols 13 and 14:} The stellar mass-to-light ratio, \mls, in solar units.
Col. 17 lists the measured \mls, and col. 18 the \mls corrected for
inclination and extinction effects.

{\bf Col 15:} The ratio of dark to light matter, $\frac{M_D}{M_L}$}, within
\RCf.

\vspace{0.15truein}
\noindent
{\bf Comments:} 

{\small
\noindent
{\it AGC110759/NGC507-A1}: ORC flux fairly faint.  \newline
{\it UGC916}: Use HI flux but not HI width (detection looks real,
but low S/N, baseline quite sloped, horns not defined so width, systemic
velocity not measurable).  \newline
{\it UGC1045}: Use TF I-band data from Han \& Mould 1992; assume 2R$_d$ lies
in long, flat region of ORC.  \newline
{\it UGC1404}: Use TF I-band data from Han \& Mould 1992; assume R$_d$ = R$_{23.5} / 3.5$.  \newline
{\it UGC1437}: ORC flux suggests strong HII regions along major axis.  \newline
{\it AGC120684}: H$\alpha$ spectrum in absorption; HI profile shows wider velocity width than ORC.  \newline
{\it AGC130014/CGCG415-058}: Use TF I-band data from Han \& Mould 1992; 
assume R$_d$ = R$_{23.5} / 3.5$.  \newline
{\it UGC2561}: Marginal HI detection (see Magri \etal 1988); calculate HI
gas mass from rms, not flux.  \newline
{\it UGC2581}: Marginal HI detection with GB300 (see Magri \etal 1988); calculate HI
gas mass from rms, not flux.  \newline
{\it UGC2617}: Treat as if no HI data (clearly confused with known companion).  \newline
{\it UGC2639}: H$\alpha$ spectrum shows both emission and absorption flux.  \newline
{\it UGC2642}: Very weak H$\alpha$ emission flux; only the nucleus can be traced.  \newline
{\it UGC2655}: ORC flux suggests strong HII regions along major axis.  \newline
{\it UGC2659}: No HI data in AGC archive; use Theureau \etal 1998 (not of
the same quality as our own reprocessing of archive data, because we cannot
make the same corrections.)  \newline
{\it AGC130204/CGCG540-115}: H$\alpha$ spectrum quite asymmetric, with right
side much fainter than left side.  H$\alpha$ absorption shows faintly beyond
emission; perhaps trough not completely filled?  \newline
{\it UGC2696}: H$\alpha$ absorption shows faintly beyond emission; perhaps
trough not completely filled?  \newline
{\it UGC2742}: Marginal and suspect HI detection, do not use (HI profile
implies much wider velocity width than ORC).  Difficult to extract $H\alpha$
velocity profile due to significant $\sin(i)$ effect (very face-on).  \newline
{\it UGC3255}: Continuum extremely strong around $H\alpha$ and [NII] emission
lines.  \newline
{\it AGC150073}: Gadzooks! -- ORC is of neighboring galaxy (do not use).    \newline
%
{\it AGC150118}: ORC extremely faint, in H$\alpha$ and in [NII].  Left side
less faint than right.  \newline
{\it UGC3291}: Assume that HI detection is interference peak and discard, as
ORC data are solid, HI line profile shows single peak, and $\Delta$cz = 4500
\kms.  Calculate HI gas mass from rms, not flux.  \newline
{\it UGC4575}: [NII] masked by [OH] emission line feature.  \newline
{\it UGC4865}: Inclination and corrected ellipticity from Dale \etal 1999.  \newline
{\it UGC5250}: ORC slightly truncated by edge of slit on left side.  \newline
{\it AGC210015/A148}: Irregular/ring galaxy.  \newline
{\it AGC210643}: Treat as nondetection in HI.  \newline
{\it UGC6697}: H$\alpha$ spectrum quite asymmetric and disturbed; all ORC emission lines 
extremely wide and strong.  \newline
{\it UGC6718}: No HI in AGC archive; use Giovanelli \& Haynes 1985 
(lesser quality than reprocessing of archive data).  \newline
{\it UGC6724}: Patchy HII along major axis.  \newline
{\it UGC6746}: Patchy HII along major axis.  \newline
{\it AGC210816/CGCG127-062}: Extremely faint H$\alpha$ flux, especially on
right, some [NII] in nucleus.  \newline
{\it AGC211029/F1319}: Treat as if no HI data (clearly confused with known
companion).  \newline
{\it UGC6928}: No HI in AGC archive; use Bottinelli \etal 1990 
(lesser quality than reprocessing of archive data).  \newline
{\it UGC7724}: H$\alpha$ and [NII] spectra very weak along disk.  \newline
{\it AGC221147/CGCG160-031}: H$\alpha$ spectrum fainter on right side then on left.  \newline
{\it UGC8069}: Treat as a nondetection in HI.  \newline
{\it UGC8096}: H$\alpha$ emission line turns into strong absorption on left side.  \newline
{\it AGC221409}: [NII] extent matches H$\alpha$ extent, except on truncated side of spectrum.  \newline
{\it UGC8161}: ORC suggests patchy HII regions along major axis.  \newline
{\it UGC8194}: Use TF I-band data from Bernstein \etal 1994; assume R$_d$ = R$_{23.5} / 3.5$.   \newline
{\it AGC250443/CGCG049-108}: Inclination and corrected ellipticity from Dale \etal 1999.  \newline
{\it AGC250472/CGCG049-127}: Inclination and corrected ellipticity from Dale \etal 1999.  \newline
{\it AGC250632/CGCG049-151}: Inclination and corrected ellipticity from Dale \etal 1999.  
Note $cz_{rad} = 9849$ \kms, $cz_{orc} = 9965$ \kms.  \newline
{\it AGC250740/CGCG077-119}: Inclination and corrected ellipticity from Dale \etal 1999.  \newline
{\it UGC9838}: No HI in AGC archive; use Freudling, Haynes \& Giovanelli 1992 
(lesser quality than reprocessing of archive data).  \newline
{\it UGC9844}: Inclination and corrected ellipticity from Dale \etal 1999.   \newline
{\it AGC250803/CGCG077-126}: Inclination and corrected ellipticity from Dale \etal 1999.   \newline
{\it AGC250896/CGCG077-135}: Inclination and corrected ellipticity from Dale \etal 1999.   \newline
{\it AGC251744/CGCG476-112}: Disregard HI data altogether (profile is
confused with AGC251498); ORC replaced with Dale version ($\Delta$PA =
$13^{\circ}$).  Identified as AGC251499 in Vogt 1995.  \newline
{\it AGC251400/CGCG108-031}: Treat as if no HI data (clearly confused with known companion).  \newline
{\it UGC10121}: Inclination and corrected ellipticity from Dale \etal 1999.   \newline
{\it AGC251424/CGCG108-037}: Inclination and corrected ellipticity from Dale \etal 1999.   \newline
{\it AGC251503}: Inclination and corrected ellipticity from Dale \etal 1999.   \newline
{\it AGC251436/CGCG108-043}: Inclination and corrected ellipticity from Dale \etal 1999. 
ORC offset in position angle from Dale spectrum.  \newline
{\it AGC251445/IC1157}: Weak H$\alpha$ absorption, weak [NII] emission.  \newline
{\it AGC251509}: Inclination and corrected ellipticity from Dale \etal 1999.   \newline
{\it UGC10131}: ORC quite faint.  \newline
{\it AGC251510}: Inclination and corrected ellipticity from Dale \etal 1999.   \newline
{\it UGC10166}: Treat as nondetection in HI.  \newline
{\it AGC260112/CGCG108-108}: ORC offset in position angle from Dale spectrum.  \newline
{\it AGC260116/CGCG108-107a}: [NII] masked by [OH] emission line feature.  \newline
{\it AGC260953}: Disk scale length from Dale \etal 1999.  H$\alpha$ on both
sides, but [NII] only on the left, [NII] masked by [OH] emission line feature.  \newline
{\it UGC10177}: No HI in AGC archive; use Giovanelli \& Haynes 1985 
(lesser quality than reprocessing of archive data).  \newline
{\it UGC10180}: No HI in AGC archive; use Giovanelli \& Haynes 1985 
(lesser quality than reprocessing of archive data).  \newline
{\it AGC260146}: Inclination and corrected ellipticity from Dale \etal 1999.   \newline
{\it UGC10190}: [NII] masked by [OH] emission line feature.  \newline
{\it UGC10192}: Identified as {\it AGC260184} in Vogt 1995.  No HI in AGC
archive; use Huchtmeier \& Richter 1989 
(lesser quality than reprocessing of archive data).  \newline
{\it UGC10193}: [NII] masked by [OH] emission line feature.  Note moderately
close companion in ORC ($\Delta$cz = 600 \kms) and in I-band.  \newline
{\it UGC10195}: No HI in AGC archive; use Giovanelli \& Haynes 1985 
(lesser quality than reprocessing of archive data).  \newline
{\it AGC260208/CGCG108-140}: Inclination and corrected ellipticity from Dale \etal 1999.   \newline
{\it AGC260226/IC1192}: Inclination and corrected ellipticity from Dale \etal 1999.   \newline
{\it AGC260245}: Inclination and corrected ellipticity from Dale \etal 1999. 
H$\alpha$ beyond [NII] is very faint on left, bright on right.  \newline
{\it AGC260246/CGCG108-155}: Inclination and corrected ellipticity from Dale \etal 1999.  \newline
{\it AGC260274/CGCG108-157}: H$\alpha$ outlying points on left are quite faint.  \newline
{\it UGC10287}: ORC offset in position angle from Dale spectrum.  \newline
{\it UGC10389}: No HI in AGC archive; use Huchtmeier \& Richter 1989 
(lesser quality than reprocessing of archive data).  \newline
{\it AGC260599}: ORC very probably has large position angle offset (do not use velocity profile).  \newline
{\it UGC10415}: Face-on galaxy, ORC is quite noisy because of inclination
angle factor, major axis was difficult to determine.  \newline
{\it AGC260929}: Faint, patchy HII along major axis.    \newline
{\it AGC260622}: Faint, patchy HII along major axis.    \newline
{\it AGC260930}: Fairly strong H$\alpha$ emission spectrum, but [NII] extremely faint.  \newline
{\it AGC260934}: [NII] masked by [OH] emission line feature.  \newline
{\it UGC10661}:Use HI flux but not width (detection looks real, but low S/N,
horns not defined).  \newline
{\it UGC11348}: No HI data in AGC archive; use Theureau \etal 1998 
(lesser quality than reprocessing of archive data).  \newline
{\it AGC330564/CGCG476-075}: Strong, solid body inner ORC profile rises
above V$_{max}$ - bar?  \newline
{\it AGC330712}: Strong H$\alpha$, but no [NII] to trace along ORC.  \newline
{\it UGC12721}: Patchy HII along disk (isolated points appear to be real, match I-band).  \newline
{\it AGC330721}: Treat as if no HI data; likely confused with known companion.  \newline
{\it AGC330768/CGCG476-112}: Sole case of strong ORC asymmetry outside of
cluster cores, from interacting neighbor at 16 \hkpc.  \newline
{\it AGC331185/F2532}: New AGC name for AGC331096 (deactivated).  HI
spectrum has been published as AGC331096, and recorded in NED.  Identified as
AGC331096 in Vogt 1995.  \newline
{\it AGC330918/CGCG498-012}: Use TF I-band data from Han \& Mould 1992; 
assume R$_d$ = R$_{23.5} / 3.5$.  \newline
{\it AGC330958/CGCG477-020}: HI marginal (S/N barely acceptable) and
$\Delta$cz = 1200 \kms from ORC; assume nondetection in HI and calculate HI
gas mass from rms, not flux.  \newline
{\it AGC331013/CGCG478-003}: HI marginal (S/N barely acceptable) and
$\Delta$cz = 200 \kms from ORC; assume nondetection in HI and calculate HI
gas mass from rms, not flux.

}
\vspace{0.15truein}

\begin{table*} [htbp]
  \caption{Galaxy Parameters III}
  {\small
  \begin{center}
  \begin{tabular} {l r r r r r r r r r r r r r r} 
  \tableline
  \tableline
  Names & 
  \multicolumn{1}{c}{$RW_0$} & \multicolumn{1}{c}{$RW_1$} & \multicolumn{1}{c}{$RW_2$} & \multicolumn{2}{c}{$M_{HI} (obs, exp)$} & 
  \multicolumn{1}{c}{$i$} & \multicolumn{1}{c}{$I$} & \multicolumn{1}{c}{$M_I$} & 
  \multicolumn{1}{c}{D/T} & \multicolumn{1}{c}{R$_d$} & \multicolumn{1}{c}{$R_{23.5}$} & 
  \multicolumn{2}{c}{$\left[\frac{M}{L}\right]$} & \multicolumn{1}{c}{$\frac{M_D}{M_L}$} \\
  & 
  \multicolumn{1}{c}{--} & \multicolumn{1}{c}{km s$^{-1}$} & \multicolumn{1}{c}{--} & 
  \multicolumn{2}{c}{$\log 10^{10} M_{\odot}$} & 
  \multicolumn{1}{c}{$^{\circ}$} & & & \multicolumn{1}{c}{\%} & \multicolumn{1}{c}{$\prime\prime$} & \multicolumn{1}{c}{$\prime\prime$} & \multicolumn{2}{c}{$\odot$} & \\
  \multicolumn{1}{c}{(1)}  & 
  \multicolumn{1}{c}{(2)}  & \multicolumn{1}{c}{(3)}  & \multicolumn{1}{c}{(4)}  & 
  \multicolumn{1}{c}{(5)}  & \multicolumn{1}{c}{(6)}  & \multicolumn{1}{c}{(7)}  & \multicolumn{1}{c}{(8)}  & \multicolumn{1}{c}{(9)}  & 
  \multicolumn{1}{c}{(10)} & \multicolumn{1}{c}{(11)} & \multicolumn{1}{c}{(12)} & \multicolumn{1}{c}{(13)} & \multicolumn{1}{c}{(14)} & 
  \multicolumn{1}{c}{(15)} \\
  \tableline 
  U00809/520-033 & 314 & 289 & 291( 6) &  9.31 &  9.25 & 83 & 13.55 & -22.30(03) & 98 & 10.32 & 49.5 & 0.64 & 2.06 & 0.15 \\
  U00841/502-025 & 290 & 262 & 266( 8) &  9.27 &  9.54 & 80 & 13.58 & -22.18(03) & 97 &  8.90 & 43.5 & 0.62 & 1.71 & 0.13 \\
  U00927/N496    & 297 & 271 & 322(12) &  9.72 &  9.61 & 57 & 12.47 & -22.93(05) & 97 & 14.77 & 52.3 & 1.07 & 1.61 & 0.08 \\
  U00940/521-016 & 288 & 260 & 331(13) &  9.57 &  9.48 & 52 & 13.63 & -22.43(04) & 93 &  8.77 & 30.5 & 1.71 & 2.42 & 0.19 \\
  U00944/N512    & 522 & 493 & 499( 7) &  9.41 &  9.36 & 81 & 12.04 & -23.70(02) & 91 &  8.70 & 55.4 & 0.50 & 1.30 & 0.28 \\
  \tableline \\
  \multispan{15}{The complete version of this table is in the electronic edition of the Journal.  The printed edition contains only a } \\
  \multispan{15}{sample.} \hidewidth
  \end{tabular}
  \end{center}
  }
  \label{tab:crdnl1c}
\end{table*}

\begin{figure*} [htbp]
  \begin{center}\epsfig{file=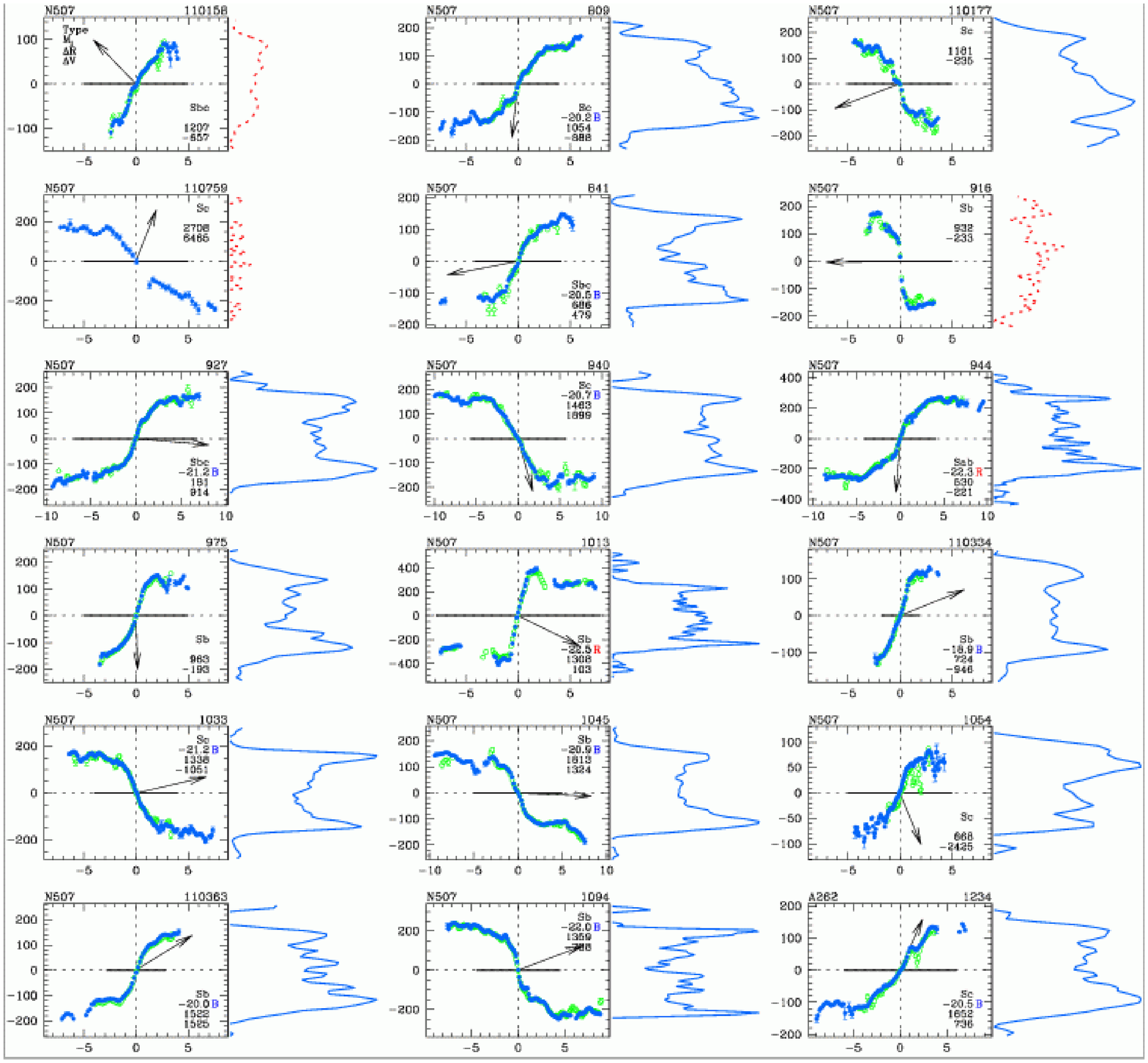,width=6.0truein}\end{center}
  \caption[Rotation Curves and 21~cm line profiles.]  
  {Rotation curves are shown for the entire sample, sorted by right ascension
  and labeled by AGC name.  Filled blue circles show \Halpha data, and open
  green circles \fnii data; error bars are plotted when larger than 10 \kms.
  The x-axis shows the radial distance along the major axis, in \hkpc,
  and the y-axis the velocity, in \kms.  The cluster name is noted for
  galaxies within two \hMpc of a cluster center, with the galaxy radius
  (\hMpc) and velocity offset ($\sigma$) inset below the morphological type.
  The absolute magnitude M$_I$ and $B-I$ color (B/R) are added if available.  A
  solid $\pm2R_d$ disk length is drawn along the major axis.  For galaxies
  within one \hMpc of a cluster center, an arrow from the center of the
  rotation curve points towards the cluster center (the angle between the
  x-axis and the arrow indicates the position of the major axis relative to
  the cluster center.)
  To the right of each rotation curve we present the \HI line profile,
  preserving the velocity scale on the y-axis.  The blue line profiles are
  scaled along the x-axis to fill the available horizontal space rather than
  in absolute flux units, but the line profile is scaled lower and drawn as a
  dotted red line for galaxies which are \HI deficient by more than a factor
  of 2.5 (log \HI$_{def} \ge$ 0.40).
  [The complete version of this figure is in the electronic edition of
  the Journal.  The printed edition contains only a sample.]}
  \label{fig:rc-fl1}
\end{figure*}

\begin{figure*} [htbp]
  \begin{center}\epsfig{file=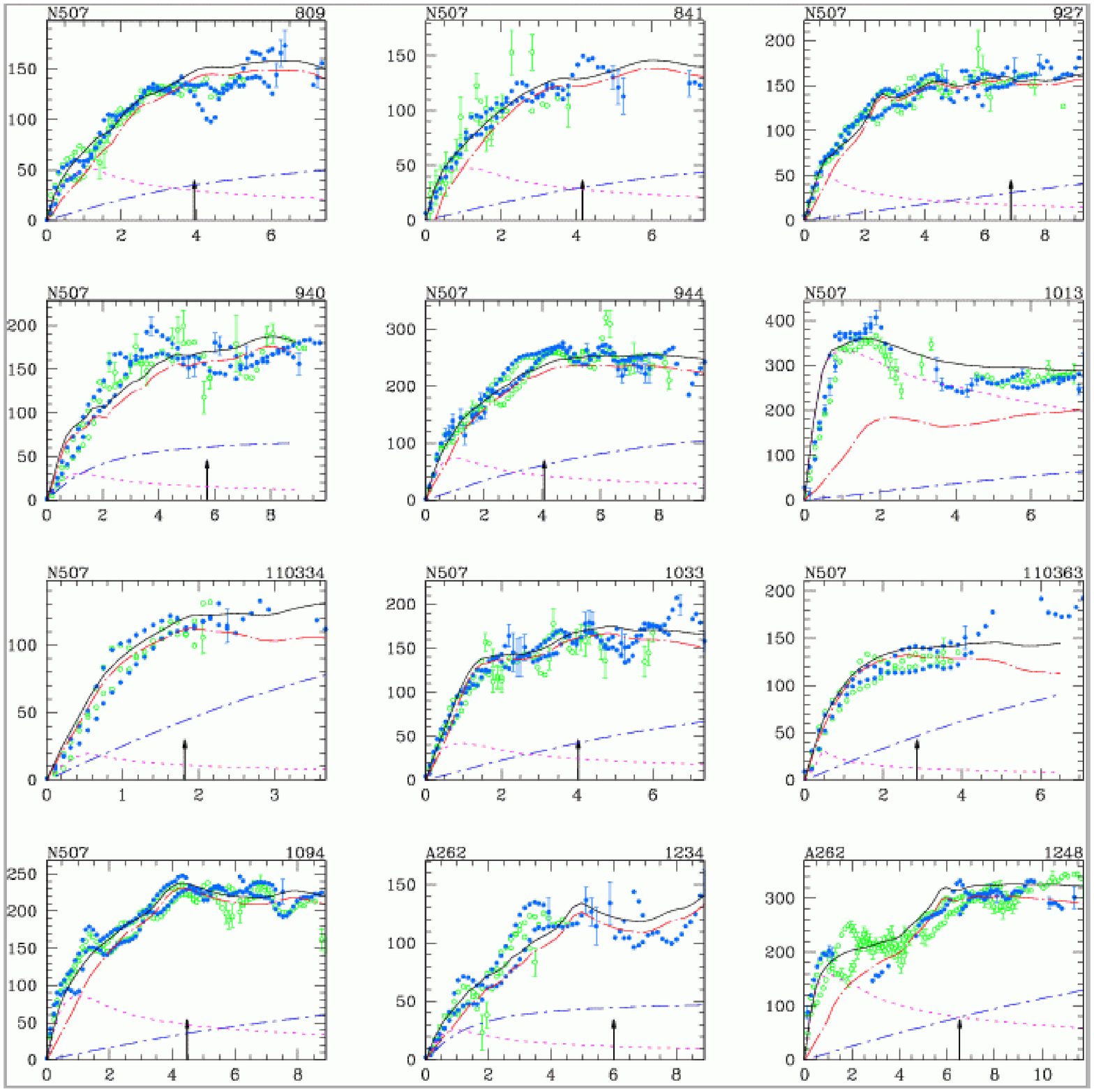,width=6.0truein}\end{center}
  \caption[Mass Models.]
  {Rotation curves are folded about their centerpoints for the galaxies for
  which we have \Ib data, sorted by right ascension and labeled by AGC name.  
  Solid blue circles designate the \Halpha data, and the open green circles 
  the \fnii data.  Error bars are again plotted when larger than 10 \kms.  The
  x-axis shows the radial distance along the major axis, in units of
  \hkpc, and the y-axis the velocity, in units of \kms.  The profiles
  have been corrected by $\sin (i)$ to edge--on amplitude.  Maximum disk mass
  models have been superimposed on top of the rotation curves; the dotted purple 
  lines represent the bulge components, the dot-long-dashed red lines the
  disk components, the dot-short-dashed blue lines the halo components, and the
  solid black lines the total velocity models.  Upright arrows on the x--axis
  mark one disk scale length.
  [The complete version of this figure is in the electronic edition of
  the Journal.  The printed edition contains only a sample.]}
  \label{fig:rc-mm1}
\end{figure*}

\begin{figure*} [htbp]
  \begin{center}\epsfig{file=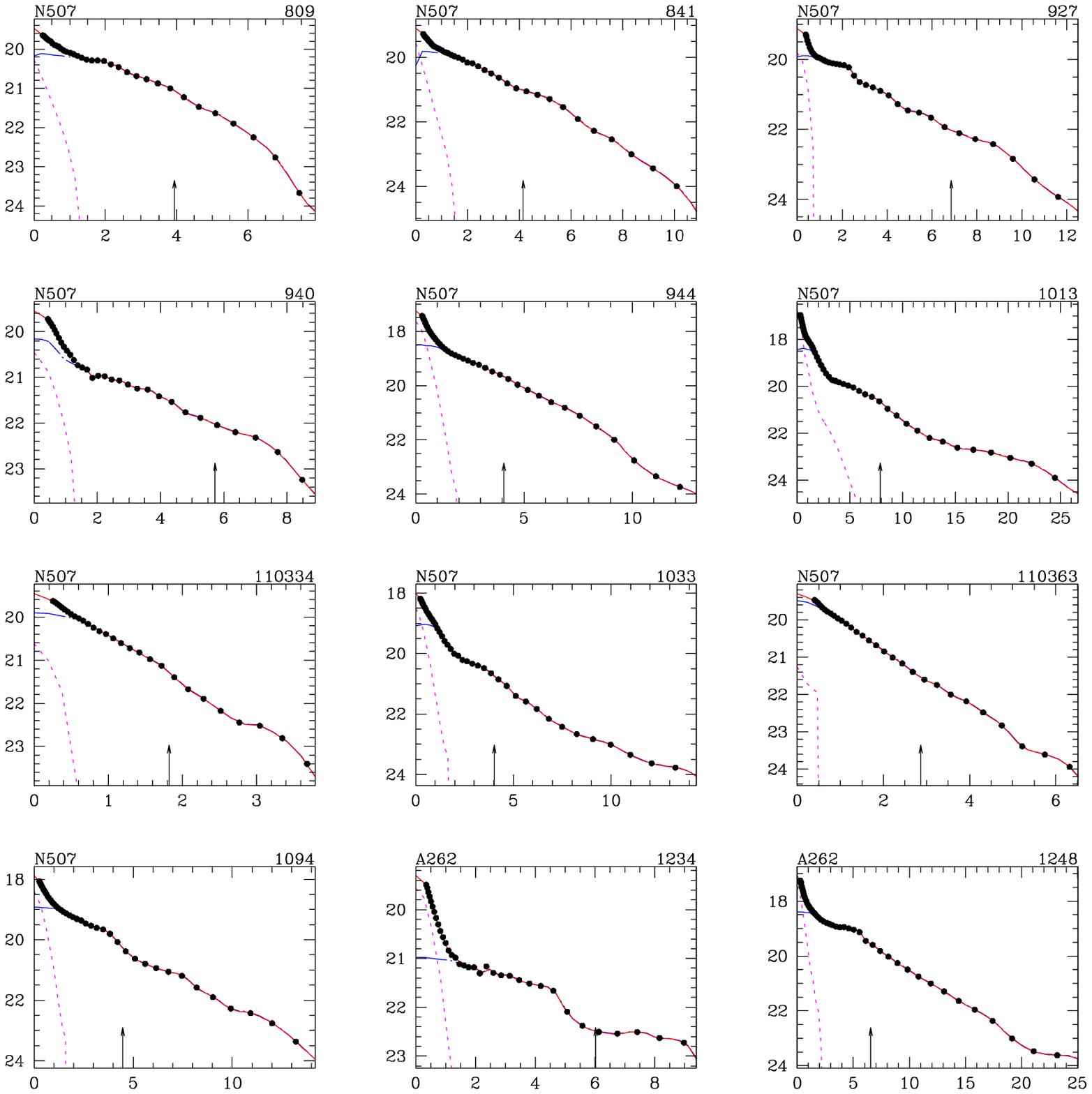,width=6.0truein}\end{center}
  \caption[Surface Brightness Profiles.]
  {\Ib surface brightness profiles are drawn for the galaxies for which we 
  have \Ib data, sorted by right ascension and labeled by AGC name.  The 
  x-axis shows the radial distance along the major axis, in \hkpc, and 
  the y-axis the surface brightness profile (red), azimuthally averaged along the 
  fitting ellipse, in magnitudes per square arcsecond.  An arrow marks 
  the radius of two disk scale lengths, and the bulge (purple) and disk (blue) 
  components have been deconvolved.
  [The complete version of this figure is in the electronic edition of
  the Journal.  The printed edition contains only a sample.] \vspace{0.11truein}}
  \label{fig:im-sb1}
\end{figure*}

Figure~\ref{fig:rc-fl1}
\footnote{Figures~\ref{fig:rc-fl1},~\ref{fig:rc-mm1},~and~\ref{fig:im-sb1} are
available online via journal records in their complete form.  Their first
pages are presented here for guidance regarding their form and content.}
presents the data on the derived rotation curves for the entire sample of
galaxies, sorted by right ascension and labeled by AGC name.  Filled circles
designate the \Halpha data, and open circles the \fnii data (where traced).
Error bars are plotted {\it only} on points for which they have an amplitude
greater than 10 \kms, to reduce obscuration of the points.  The x-axis shows
the radial distance along the major axis in units of \hkpc, and the
y-axis the velocity in units of \kms.  The cluster name is added for galaxies
within 2 \hMpc of a cluster center, and the galaxy radius (\hMpc) and velocity
offset ($\sigma$) are below.  The absolute magnitude M$_I$ is then added for
all galaxies as available.  For galaxies within 1
\hMpc of a cluster center, an arrow at the center of the rotation 
curve points in the direction of the cluster center.  (The angle between the
x-axis and the arrow indicates the position of the major axis relative to the
cluster center, the angle between the rotation curve, showing the velocity
along the major axis, and the arrow means nothing.)

To the right of each rotation curve we present the \HI line profile,
preserving the velocity scale on the y-axis to allow a comparison of the \HI
flux distribution with the rotation curve as a function of radius.  The line
profile is scaled along the x-axis to fill the available space rather than in
absolute flux units, but the line profile is drawn with a dotted line for
galaxies which are \HI deficient by more than a factor of 2.5 (log
\HI$_{def} \ge$ 0.40).

Figure~\ref{fig:rc-mm1}{$^4$} shows the rotation curves folded about their
centerpoints for the galaxies for which we have \Ib data, sorted by right 
ascension and labeled by AGC name.  Circles designate the \Halpha data, and 
triangular points the \fnii data, while filled points and empty points fall 
respectively on the approaching and receding sides of the source.  Error bars 
are again plotted {\it only} on points for which they have an amplitude greater 
than 10 \kms, to reduce obscuration of the points.  The x-axis shows the radial 
distance along the major axis in units of \hkpc, and the y-axis the 
velocity in units of \kms.  For sources where an image is available and thus 
an inclination angle is known, the profiles have been corrected by $\sin (i)$ 
to edge--on amplitude.  Maximum disk mass models have been superimposed on top 
of the rotation curves for sources for which images are available.  The dotted 
lines represent the bulge components, the dot-long-dashed lines the disk 
components, the dot-short-dashed lines the halo components, and the solid 
lines the total velocity models.  The component profiles have been terminated 
at the end of the luminosity profile for the few sources in which the optical 
rotation curve extends further out than the traced luminosity profile.  The 
upright arrows placed upon the x--axis mark one disk scale length.

Figure~\ref{fig:im-sb1}{$^4$} presents the data on the \Ib surface brightness
profiles for the galaxies within the rotation curve sample for which we have
such data, sorted by right ascension and labeled by AGC name.  The x-axis
shows the radial distance along the major axis in \hkpc, and the y-axis
the surface brightness profile, azimuthally averaged along the fitting
ellipse, in magnitudes per square arcsecond.  A an arrow marks the radius of
two disk scale lengths, and the bulge and disk components have been
deconvolved.

\section{Summary}

This study concerns the effects of the cluster environment upon the
distribution of mass and light in spiral galaxies.  We have observed a set of
329 nearby (redshift $z < 0.045$) galaxies, (1) to study the dependence of
mass--to--light distributions upon cluster environment, (2) to evaluate the
global properties of optical rotation curves (spatially extended velocity
profiles), and (3) to determine the robustness of the Tully--Fisher relation
within differing environments.  The 16 Abell clusters cover a wide range of
cluster richness, density, and luminosity, and are supplemented by the less
rich groups Cancer and \NGC 507, and 30 isolated field galaxies.  We have
examined the distribution of mass and light within the target galaxies through
a study of optical emission lines, the total \HI gas, and optical photometry.
Optical spectroscopy were taken with the 200-inch Hale Telescope and provide
separately resolved \Halpha and \fnii major axis rotation curves for the
complete set of galaxies, which are analyzed to yield velocity widths and
velocity profile shapes and gradients.  \HI line profiles provide an
independent velocity width measurement and a measure of \HI gas deficiency.
\Ib images are used to deconvolve profiles into disk and bulge components and
to determine global luminosities and inclination.  This paper is the first in
a series of three linked papers concerning this data set.  The present paper
(\pone) covers the observational data collected; the remaining two (Vogt \etal
2003; \ptwo\ and \pthree) contain the results of the analysis of cluster
properties and environmental effects.

\section{Acknowledgments}

The data presented in this paper are based upon observations carried out at
the Arecibo Observatory, which is part of the National Astronomy and
Ionosphere Center (NAIC), at Green Bank, which is part of the National Radio
Astronomy Observatory (NRAO), at the Kitt Peak National Observatory (KPNO),
the Palomar Observatory (PO), and the Michigan--Dartmouth--MIT Observatory
(MDM). NAIC is operated by Cornell University, NRAO by Associated
Universities, inc., KPNO and CTIO by Associated Universities for Research in
Astronomy, all under cooperative agreements with the National Science
Foundation. The MDM Observatory is jointly operated by the University of
Michigan, Dartmouth College and the Massachusetts Institute of Technology on
Kitt Peak mountain, Arizona. The Hale telescope at the PO is operated by the
California Institute of Technology under a cooperative agreement with Cornell
University and the Jet Propulsion Laboratory.  We would like to thank the
staff members at these observatories who so tirelessly dedicated their time to
insure the success of our observations.  We also thank Sc project team members
John Salzer, Gary Wegner, Wolfram Freudling, Luiz da Costa, and Pierre
Chamaraux, and also Shoko Sakai and Marco Scodeggio, for sharing their data in
advance of publication.  We thank the anonymous referee for helpful comments on 
the manuscript.

N.P.V. is a Guest User, Canadian Astronomy Data Center, which is operated by
the Dominion Astrophysical Observatory for the National Research Council of
Canada's Herzberg Institute of Astrophysics.  This research has made use of
the NASA/IPAC Extragalactic Database (NED) which is operated by the Jet
Propulsion Laboratory, California Institute of Technology, under contract with
NASA.  This research was supported by NSF grants AST92--18038 and AST95--28860
to M.P.H. and T.H., AST90--23450 to M.P.H., AST94--20505 to R.G., and
NSF--0123690 via the ADVANCE Institutional Transformation Program at NMSU, and
NASA grants GO-07883.01-96A to N.P.V. and NAS5--1661 to the WFPC1 IDT.

\end{document}